%% file: main.tex
\documentclass[doc,babel,american,11pt]{apa7}
\usepackage[style=apa,backend=biber,natbib=true]{biblatex}
\usepackage{csquotes}
\usepackage{mathptmx}
\usepackage{array}
\usepackage{textcomp}
\usepackage{rotating}
\usepackage{tablefootnote}
\usepackage{amsmath}
\usepackage{amssymb}
\usepackage{graphicx}
\usepackage{color}
\hypersetup{breaklinks=false,hidelinks}
\usepackage{nameref} 
\usepackage{eurosym} 
\usepackage{booktabs} 
\usepackage{subfig} 
\usepackage{float} 
\usepackage[ampersand]{easylist} 
\usepackage[textsize=small, textwidth=3.2cm, colorinlistoftodos]{todonotes} 
\usepackage{lineno}
\usepackage[graphicx]{realboxes}

\usepackage[doublespacing]{setspace}

\renewcommand{\enquote}[1]{``#1''} 

\title{\bf A dynamical scan path model for task-dependence during \\[1mm] scene viewing}
\shorttitle{A task-dependent scan path model}
\leftheader{Schwetlick et al.}
\author{Lisa Schwetlick, Daniel Backhaus, Ralf Engbert}
\authorsaffiliations{{Department of Psychology, University of Potsdam}}

\authornote{
\addORCIDlink{Lisa Schwetlick}{0000-0003-3356-8324}, \addORCIDlink{Daniel Backhaus}{0000-0002-1291-8762}, \addORCIDlink{Ralf Engbert}{0000-0002-2909-5811}

Our study is preregistered (\url{https://osf.io/dsyt2/});  all research data including computer code for simulation and analyses reported in this article is available online via the Open Science Framework (\url{https://osf.io/dsyt2/}). The manuscript was uploaded as a preprint to ArXiv (\url{https://arxiv.org/abs/2112.11067}).  Experimental data in this manuscript were previously analyzed and discussed in \citep{Backhaus2019}. 

This work was funded by Deutsche Forschungsgemeinschaft via Collaborative Research Center (SFB) 1294, projects B03 and B05 (project no.~318763901). We acknowledge additional support from grant and EN 471/16-1 by Deutsche Forschungsgemeinschaft. 

Correspondence should be addressed to Lisa Schwetlick (email: lisa.schwetlick@uni-potsdam.de).
}

\makeatletter

\begin{document}
\maketitle

{\vspace{2cm}
\centering\huge\textcolor{red}{Note: This is a pre-print. The final paper is in press and will appear in Psychological Review.}
\vspace{2cm}}

\newpage

{\Large \begin{center}
    Abstract
\end{center} }

In real-world scene perception human observers generate sequences of fixations to move image patches into the high-acuity center of the visual field. Models of visual attention developed over the last 25 years aim to predict two-dimensional probabilities of gaze positions for a given image via saliency maps. Recently, progress has been made on models for the generation of scan paths under the constraints of saliency as well as attentional and oculomotor restrictions. Experimental research demonstrated that task constraints can have a strong impact on viewing behavior. Here we propose a scan path model for both fixation positions and fixation durations, which includes influences of task instructions and interindividual differences. Based on an eye-movement experiment with four different task conditions, we estimated model parameters for each individual observer and task condition using a fully Bayesian dynamical modeling framework using a joint spatial-temporal likelihood approach with sequential estimation. Resulting parameter values demonstrate that model properties such as the attentional span are adjusted to task requirements. Posterior predictive checks indicate that our dynamical model can reproduce task differences in scan path statistics across individual observers.
\vspace{1cm}


\newpage

From the early days of eye movement research into the present, the question of how task influences the decisions on and the order of fixation locations has been of central interest. One of the first eye movement studies, the seminal but anecdotal work by \citet{Yarbus1967} suggests qualitative differences in scan paths when looking at the same image under different task instructions. Yarbus concluded that both the fixation density and sequences of fixated locations (i.e., scan paths) sensitively depend on task requirements. Within the large body of subsequent work (see below) on this topic, a variety of methods for investigating the role of task for active vision \citep{Findlay2003} have been proposed. Comparisons of eye movement measures demonstrate that spatial fixation locations as well as fixation durations are influenced by task \citep{Castelhano2009}. It was also noted that differences between tasks can be larger than the interindividual differences between observers \citep{DeAngelus2009}. 

In this paper we study a theoretical model to investigate the research question of how task demands modulate scan path generation. Modeling scan path generation provides crucial constraints on underlying cognitive, attentional, and motor processes \citep[e.g.,][]{LeMeur2015,Engbert2015,Tatler2017,Schwetlick2020}. Here we develop and analyze a mathematical model of scan path generation across tasks. First, we advance our earlier dynamical model \citep{Schwetlick2020,Engbert2015} to include the control of fixation durations in addition to fixation positions. Second, the model will be fitted to experimental scan paths from individual observers using Bayesian inference for dynamical models \citep{Schuett2017,Engbert2022}. Third, with this detailed account of scan path generation, we model task-dependence across four different viewing conditions from an earlier experimental study \citep{Backhaus2019}.

\subsection{Task differences in scene viewing}
\label{sec:task}
Early reports by \citet{Buswell1935} and \citet{Yarbus1967} lend support to the idea that eye movement patterns depend on the viewer's instruction and not just on image content and features. Such effects of top-down task impact on viewing behavior have been replicated by follow-up experimental studies \citep{Castelhano2009,DeAngelus2009,Mills2011}. Related work included investigations of eye movements during everyday tasks like preparing a cup of tea \citep{Land1999} or a sandwich \citep{Hayhoe2003}. During such tasks, gaze control supports general motor control by either moving relevant information to the central visual field \citep{Ballard1997,Land2009} or by selecting object information needed later during the task to prepare future movements \citep{Pelz2001}. Important examples include driving \citep{Land2001}, cycling \citep{Vansteenkiste2014}, walking \citep{Matthis2018,Rothkopf2007}, and ball games \citep{Land2000,Land1997}. These experimental designs move away from the typical lab-based scene viewing paradigm and contribute to a more ecologically valid account of eye guidance. In the typical scene viewing paradigm where no task is given, participants are free to choose their own objective or task, which is hidden from the researcher's access \citep{Tatler2011}. Given the relevance of specific viewing strategies to different tasks, scene viewing without clear task instruction might thus be difficult to interpret. 

The rise of modern machine learning techniques motivated purely data-driven research on scan path patterns to identify task from experimental fixation sequences. Initially, work on this topic generated mixed results \citep[see][for a detailed review]{Boisvert2016}. Based on scan path visualization of their underlying data, \citet{Greene2012} found that neither human experts nor any of three proposed pattern classifiers were able to reliably infer which task the observer was performing. The same experimental data were later reanalyzed by \citet{Borji2014}. As a result, the classifier could be improved significantly, showing  35\% accuracy for a four-task classification data set, where the reanalysis included more spatial data in the form of low resolution fixation density patterns. Performance was further boosted by accounting for inter-individual and image differences \citep{Kanan2014}. Furthermore, a classifier trained using a hidden Markov model approach indicated that additional diagnostic information for successful task prediction is contained in the scan path dynamics \citep{HajiAbolhassani2014}. 

Experimental results agree that the given task significantly affects gaze characteristics. Specifically, \citet{Castelhano2009} found that both number of fixations and fixation durations varied with task and that fixated areas were qualitatively different between tasks. Other studies also found effects of task on temporal (e.g., fixation duration) as well as spatial (e.g., saccade amplitude) measures \citep{Mills2011,Bonev2013}. Search tasks have also been found to lead to an extended range of fixation locations compared to free viewing material \citep{Tatler2007}. However, finding systematic differences for the type of task, such as free viewing or search has yielded inconsistent results. While \citet{Mills2011} found shorter fixation durations for search tasks compared to free viewing, results disagree about saccade amplitude with more recent findings by \citet{Backhaus2019}. Because of the variety of tasks and stimuli in experimental paradigms, however, it can be expected that comparisons of results across studies is not straightforward and do not always lead to full agreement.

Taken together, experimental work as well as machine-learning classification paint a consistent picture that individual differences, spatial selection, and aggregate eye movement measures are specific for tasks. Classification success depends highly on the particular set of features selected and on the type of classification algorithm used. A number of studies used features of varying abstraction, from the very basic saccade amplitude and fixation duration to global or local image features \citep{Boisvert2016} or transition probabilities between identified regions of interest \citep{Coutrot2017}. In the next sections we discuss the role of process-oriented models in scene-viewing research, with a particular focus on scan path generation.

\subsection{Theoretical models of visual attention during scene viewing}
Human eye movements in natural scene viewing are guided by visual attention \citep{Itti2001}, which is modulated by image-dependent features. Basic research showed that saccadic eye movements follow the locus of attention \citep{Deubel1996,Kowler1995}. This tight coupling of attention and saccades is exploited in experimental work, where gaze positions are typically equated with the locus of visual attention \citep{Henderson2003}. It should be noted, however, that there are pronounced deviations between visual attention and gaze position around the time of saccade \citep{Deubel1996,Kowler1995}. For example, we recently showed in a mathematical modeling study that effects of perisaccadic attention can explain effects of saccade statistics in scene viewing \citep{Schwetlick2020} within the SceneWalk model \citep[][see below]{Engbert2015}. The following section discusses primarily image-computable models for the spatial, time-averaged distribution of fixation positions as a proxy for visual attention, which constitute a large part of the literature on the topic.

Past modeling work shows that image-dependent influences contribute strongly to predictions of the overall gaze positions when viewing natural scenes. Relevant image features include local luminance contrast and edge density \citep{Mannan1997, Reinagel1999, Tatler2005, Parkhurst2003}. Much of the early modeling work to predict fixations involved the detection of these features in images to generate an activation map, in which high activation represents areas rich in features relevant for the viewing task. 

In 1985, \citeauthor{Koch1985} proposed a model that (i) simultaneously computes various feature maps for color, intensity, edges, and feature popout detected by biologically plausible components such as center-surround receptive fields and that (ii) integrates these maps into a master map called {\sl saliency map}. This idea was later implemented as a computational model \citep[see][, for an overview]{Itti2001}. More feature-based saliency models followed, adding other spatial features or statistics (e.g., \cite{Harel2007, Torralba2006, Harel2007, Bruce2009, Tatler2006}). 

One way of evaluating saliency models is to compare the generated saliency maps to empirical fixation densities from eye movement experiments \citep{Parkhurst2002, Tatler2005}. Linking the two concepts is the assumption that attention and eye movement are closely related \citep{Henderson2003}, as discussed above. In order to be able to compare a two-dimensional density and a series of fixation locations, metrics of different sophistication have been proposed \citep{Kuemmerer2015}, including AUC score \citep{Tatler2005} and Kulback-Leibler divergence \citep{LeMeur2015}. The popular MIT/Tübingen Saliency Benchmark\footnote{See \href{https://saliency.tuebingen.ai}{https://saliency.tuebingen.ai}} employs seven such metrics, each emphasizing different aspects and yielding different results. More recently, \cite{Kuemmerer2015} suggested to use the concept of information gain based on statistical model likelihood applied to saliency modeling, as a statistically well-founded alternative to ad-hoc metrics.

Beyond simple feature-based attention, processes of target selection can also include higher level concepts such as objects or contextual guidance \citep{Torralba2006, Nuthmann2010}.  A long-standing debate exists between the more top-down interpretation that emphasizes cognitive relevance and meaning \citep{Henderson2007, Henderson2009, Henderson2019, Henderson2017} and the traditional saliency oriented approach \citep{Pedziwiatr2019}. The distinction is not as clear-cut as some discussions suggest, however, as clusters of low-level image features are also indicative of objects. Research does show that incorporating information about object locations into saliency models makes them more accurate \citep{Kuemmerer2014a}. Thus, eye movements are driven both by basic features detected by low-level vision as well as more advanced levels of cognitive processing \citep{Schuett2018}.

Using deep learning techniques and neural network architectures, recent models of general visual saliency achieved considerably improved prediction accuracy \citep{Bylinskii2015a,Kuemmerer2017} compared to earlier approaches. Neural network models are trained using experimental fixation data, which represent image-driven as well as meaning- and task-dependent influences. Correspondingly, although the outcomes are referred to as saliency maps, models trained using this data-driven approach produce descriptions of eye movements which can no longer be dissociated into particular layers of cognition. 

It is also worth noting that almost all experimental data used in this data-driven approach are acquired from participants who viewed pictures without a specific task. The underlying assumption is that simple picture viewing results in the most natural behavior. This was criticized by \citet{Tatler2011}: \enquote{It seems more likely that free viewing tasks simply give the subject free license to select his or her own internal agendas} (p. 4).

Besides the viewing task, another aspect of ecologically valid, real-world conditions is the possibility of body movement \citep{Backhaus2019, Matthis2018}. In order to evaluate whether typical lab restrictions, e.g., head stabilization on a chin rest, limit the generalizability of results \citep{Tatler2011}. \citet{Backhaus2019} suggested the use of mobile eye-tracking which permitted natural posture and postural fluctuations \citep{Collins1995}, with the general finding that task effects are robust with respect to changes in body posture while viewing images. 

Modeling of visual attention during scene viewing often focuses exclusively on the spatial aspect of \emph{where} fixations are placed on an image. Next, we discuss how time-ordered fixation sequences are generated and which factors influence scan path dynamics via biologically inspired mechanisms---including the timing of saccades \citep{Henderson2003}.

\subsection{Biologically inspired models of scan path generation}
\label{sec:biological_modeling}
During active vision \citep{Findlay2003}, our gaze continually explores the visual environment by producing saccadic movements. \citet{Findlay1999} proposed an influential conceptual model for generating saccades, which claims validity for a variety of experimental paradigms and situations. A fundamental assumption in the model is that two partially separate pathways exist for temporal and spatial control processes. Both pathways are composed of a hierarchy of levels from automatic to higher-level cognitive control, where each level has high biological plausibility. This basic architecture turned out to be successful in a variety of cognitive tasks, such as reading \citep{Engbert2002,Engbert2005,Seelig2020,Rabe2021} and scene viewing \citep{Engbert2015,Schwetlick2020}. Thus, there is theoretical support for the existence of separate pathways for to spatial and temporal control of gaze.

As discussed in the previous section, significant progress has been made on models that predict the spatial control of gaze position, often termed visual saliency modeling \citep{Itti2001}. These models aim at predicting the 2D density of fixations on a given scene. From the beginnings of the research tradition, biological plausibility played an important role for the development of these models \citep{Koch1985}. However, initially there was little interest in making use of saccade statistics that were more detailed than the spatial density of fixations.

While the role of scan paths \citep{Yarbus1967} and sequential effects in sequences of saccades \citep{Noton1971a,Noton1971b} was noted early on, it was only much later that the dissatisfaction with purely saliency-based models stimulated interest in scan path generation \citep{Zelinsky2008}. For example, \citet{Tatler2009} demonstrated that adding oculomotor principles to models could significantly improve the predictive power in spatial selection. Thus, effects of the previous fixation location on the selection of the upcoming gaze position were identified as an important modeling goal \citep{LeMeur2015} for understanding principles of human gaze control.

The success of mathematical models to reproduce human gaze positions stimulated interest in theoretical models for fixation durations \citep{Henderson2003}. Since each fixation is bounded by two saccades, effects of oculomotor preparation and execution are highly relevant to the statistics of fixation durations and saccade timing. \citet{Nuthmann2010} proposed a random-walk model for timing of saccades (CRISP) to explain data from an experimental paradigm with delayed scene onsets. Similarly, \citet{Tatler2017} used a activation-based rise-to-threshold unit \citep{Reddi2000} for the generation of saccadic onset-times in their model (LATEST). The LATEST model \citep{Tatler2017} is a combined model of spatial and temporal control. This approach represents of mixture of process-oriented \citep[LATER unit; ][]{Reddi2000} and data-driven modeling (spatial aspects). As a result, the LATEST model demonstrates important effects on the integration of spatial and temporal control of saccades. However, the data-driven components of LATEST offer limited insight into the biological processes that generate the behavior. 

A recent paper that contributed to the conceptual advancement in the field of eye-movement modeling was published by \citet{Kucharsky2021}. In this theoretical study, a model for fixation durations from the broad class of information accumulation or drift-diffusion models \citep{Smith2004} was extended by a spatial component. The integrated model termed WALD-EM was successful in modeling many aspects of saccade statistics and distributions of fixation durations. Most important for theory building, a combined spatiotemporal likelihood function was used, which provides the basis for rigorous statistical inference.

Statistical, functional, and mechanistic modeling in cognitive science take on vastly different roles \citep{Bechtel2010}. While statistical models are mainly descriptive, functional and mechanistic models are process-oriented and propose specific interactions between different subsystems. Thus, specific assumptions can be tested against experimental data, so that the plausibility of biologically-inspired mechanisms can be tested \citep{Engbert2021}. The more grounded in experimental evidence and the more mechanistic the model, the more compelling are the conclusions are about the explanation of an observed effect \citep{Bechtel2010}. Moreover, in mechanistic, generative models it is possible to interpret the model parameters with respect to the processes in the visual, attentional, and oculomotor systems. Within this class of models we developed the SceneWalk model \citep{Engbert2015,Schuett2017,Schwetlick2020}, which is in agreement with the framework proposed by \citet{Findlay1999}. 

In the SceneWalk model, fixation selection is based on a time-dependent priority map that is influenced by the current gaze position, time-independent fixation density, and previously fixated locations. More recently, \citet{Schwetlick2020} added perisaccadic attentional processes that improved the model's performance with respect to a variety of scan path metrics including complex effects such as modulations of the mean fixation durations by saccade turning angle. With respect to task influences discussed in the current work, we expect that differences in scan paths across tasks will be reflected in differences of the numerical values of model parameters across tasks. The ability of dynamical process-oriented models to reproduce  differences in behavior via parameter adaptation supports the underlying by demonstrating generalizablity. Our assumptions of how parameter values vary between tasks is based on prior experimental work showing that repeated viewing of the same natural scenes induce differences in saccade statistics (e.g., distribution of saccade lengths). These differences are compatible, in the model, with a smaller perceptual span during second viewing of the same image compared to the first viewing \citep{Trukenbrod2019}.

\subsection{The role of saliency for dynamics }
\label{sec:saliency}
Due to intense research, the scientific literature on modeling of static saliency (2D fixation density) has grown enormously, while scan path modeling is a comparatively new field of quantitative modeling. In this paper we use the SceneWalk model of eye movement dynamics \citep{Engbert2015,Schwetlick2020} investigate the principles of task-dependent scan paths. In order to generate scan paths, the model relies on activation maps which approximate visual saliency. As a stable upper bound, in earlier studies we used experimental fixation densities. Alternatively, the model could also be combined with a saliency model and generate eye movements from computer-generated saliency maps.

The aim of the current study is to investigate the modulation of underlying processes of static and dynamic components of eye guidance caused by task variation. It is important to note that the term {\sl saliency map} has been used to describe different concepts in the literature.  As discussed above, visual saliency initially referred to very low-level image features like edges \citep{Itti2001}. This early concept of saliency is completely devoid of higher level influences such as task. Later, however, the concept of saliency was expanded to include all influences that improve predictions on visual attention as indicated by gaze \citep{Kuemmerer2014a}. It is clear that if complex models are fitted to empirical fixation densities, then higher-level factors such as task are difficult to separate from low-level vision. Recently, there has been much discussion about the importance of high level image features versus meaning as a main predictor of eye movements \citep{Henderson2019,Henderson2018,Henderson2017,Pedziwiatr2021,Pedziwiatr2019}. While this discussion is clearly relevant to the distinction between top-down and bottom-up influences in vision \citep{Schuett2018}, the focus of the current work is on the interaction between saliency, task, and the dynamics of scan path generation.

As discussed above there is empirical evidence that viewing strategies during scene viewing depend on the given task. One possibility is that the main cause for this difference is an adjustment of the prioritization of visual information. Based on this assumption, elements in the visual display are weighted by attention according to their importance to the task. This hypothesis requires the input saliency for the eye movement model to be separate \textit{task-specific saliency} maps for each image and task. As an alternative hypothesis we might consider differences between tasks to be attributed to the tuning of saccade dynamics to particular tasks. Here, task-specific weighting of image features can be neglected and the eye movement model uses the same \textit{general purpose saliency} input per image for all tasks.  To represent this idea, we use a general fixation density from a free viewing task as a basis for task-specific model parameter estimation and validation. It seems likely that task-specific saliency effects as well as task-specific eye movement effects will play a role.  

Here we present the results for two alternative models using general and using task-specific fixation densities as input to the eye movement model. We then compare the resulting performance of both models. The general fixation density in the model is related to the task-independent interpretation of saliency. The saliency map that is passed into the model is constructed using experimentally recorded scan paths from a separate free-viewing experiment using the same images \citep{Backhaus2022}. The task-specific saliency version of the model is identical, with the difference that separate fixation densities were constructed using gaze data from only one of four different task conditions. 

With concurrent models for the same experimental data, model inference has become an increasingly important topic in cognitive modeling. Recently, the numerical tools and the computation power have become available to carry out rigorous parameter inference and model comparisons \citep{Schuett2017,Schwetlick2020}, in particular, if the likelihood function for the model can be computed or approximated. In the next section, we discuss statistical inference for dynamical models.

\subsection{Bayesian parameter inference for dynamical models}
\label{sec:inference}
Dynamical models of eye-movement control generate specific predictions for sequential dependencies of fixations over time. As a consequence, the full potential of statistical inference for dynamical models unfolds if model predictions are evaluated based on fixation sequences \citep{Engbert2022}. This approach requires sequential predictions for upcoming fixations, advanced computational methods, and sufficient computing time \citep{Schuett2017}. Many state-of-the-art methods for parameter inference in cognitive models, however, are based on ad-hoc performance metrics \citep{LeMeur2015,Tatler2017,Engbert2015,Zhou2021}. Often (but not always), such ad-hoc metrics ignore the sequential structure of scan paths. In these cases, researchers choose relevant metrics and compute a loss function that indicates how closely simulated data resemble experimental data based on the pre-defined metrics. Model parameters are obtained by optimization of the loss function when model parameters are varied. As a result, model inference is subjective (i.e., dependent on the choice of the  loss function) and difficult to generalize, since arbitrary metrics will optimize the model to reproduce some aspects of the model while ignoring others.

A statistically well-founded alternative is based on the likelihood function $L_M(\theta|\mbox{data})$ of a model $M$ with parameters $\theta$ given an experimental data set \citep{Myung2003}. The likelihood is defined as the conditional probability $P_M$ for observing the data in the context of model $M$ specified by parameters $\theta$, i.e.,
\begin{equation}
\label{eq_deflikelihood}
L_M(\theta|{\rm data}) = P_M({\rm data}|\theta) \;.
\end{equation}
If numerical computation of the likelihood, Eq.~(\ref{eq_deflikelihood}) is possible, then  rigorous statistical inference on model parameters and comparisons between different models are also possible, including Bayesian inference \citep{Gelman2013}. In static saliency modeling, the use of the likelihood \citep{Kuemmerer2015} is straightforward:  the saliency map is interpreted as a fixation probability and  the probability of each experimentally observed fixation position is evaluated on this probability map. In dynamical scan path modeling the process is more elaborate as explained in the following \citep{Engbert2022,Schuett2017}.

A fixation $f_i$ in a scan path ${\cal F}=\{f_1,f_2,f_3,...,f_N\}$ is given by its spatial position $(x_i,y_i)$ and its fixation duration $T_i$. Thus, a fixation $f_i=(x_i,y_i,T_i)$  is a 3-tuple. Because of the sequential nature of the scan path, the likelihood can be decomposed into a product of conditional probabilities, i.e.,
\begin{eqnarray}
\label{eq:sequenlik}
L_M(\theta|{\cal F})&=&L_M(\theta|f_1,f_2,f_3,...,f_N) \\
&=& P_M(f_1|\theta)\prod_{i=2}^N P_M(f_i|f_1,f_2,...,f_{i-1};\theta) \;,
\end{eqnarray}
where the generative model is used to estimate the probability $P_M(f_i|f_1,f_2,...,f_{i-1};\theta)$ of the $i$th fixation when enforcing the previous fixations $f_1,f_2,...,f_{i-1}$ and $P_M(f_1|\theta)$ is the first fixation that is typically known and experimentally controlled \citep{Schuett2017,Seelig2020}, so that $P_M(f_1|\theta)=1$.

In Bayesian inference, we specify a prior probability $P(\theta)$ over the model parameters and use the likelihood $L_M(\theta|{\cal F})$ to compute the posterior probability $P(\theta|{\cal F})$ using Bayes' theorem, 
\begin{equation}
\label{eq_Bayes}
P(\theta|{\cal F}) = \frac{L_M(\theta|{\cal F})P(\theta)}{\int_\Omega L_M(\theta|{\cal F})P(\theta){\rm d}\theta} \;.
\end{equation}
The integral in the denominator in Eq.~(\ref{eq_Bayes}) is typically intractable for realistic cognitive models. Therefore, the posterior probability $P(\theta|{\cal F})$ is estimated numerically via Markov Chain Monte Carlo methods \citep{Gilks1995}.  We will discuss the specific numerical procedures for parameter estimation in the methods sections.

\subsection{The current study}
The research goal of this study was to carry out model-based analyses of task effects on scan path generation. The starting point for our modeling work will be the SceneWalk model  for scan path generation during scene viewing \citep{Schwetlick2020,Schuett2017,Engbert2015}. Recently, we included peri-saccadic attentional effects, which reproduced correlations between saccade turning angles with saccade lengths and fixation durations \citep{Schwetlick2020}. This variation of the SceneWalk model can reproduce systematic variations in mean fixation durations. However, the peri-saccadic principles do not represent an explicit timing mechanism for saccades as proposed by \citet{Nuthmann2010} \citep[see also][]{Laubrock2013,Tatler2017}.

Since explicit timing effects can be expected in task-dependent scene viewing, here we developed a further version of the model that includes a mechanistic timer. As shown by the LATEST model \citep{Tatler2017}, the saliency value at fixation exerts a negative effect on the decision rate, which translates into prolonged mean fixation durations for fixation location with higher saliencies compared to lower saliences. In order to investigate a coupling between temporal and spatial information in the SceneWalk model \citep{Schwetlick2020}, we introduce a timing mechanism that enables the local saliency to influence mean fixation durations. This addition combines fixation durations and fixation locations into one coherent model, improving our general framework for generating fixation sequences. Coupling parameters for the spatial and temporal components are estimated from experimental data.

The structure of the manuscript is as follows. We start with a detailed explanation of the SceneWalk model and its underlying activation dynamics with activation and inhibition pathways \citep{Schwetlick2020}. Next, we extend the model to include the explicit timing mechanism for saccade generation. The likelihood function of the extended model can be decomposed into a spatial and a temporal component. We discuss the specific approach for numerical Bayesian inference. After the introduction to the model, we describe the experiment on natural scene-viewing which included a task manipulation \citep{Backhaus2019}. The Results show parameter estimation and posterior predictive checks that indicate the goodness-of-fit for various dependent variables. Based on the posterior estimates of the model parameters, we run a statistical analysis across participants and tasks that highlight how the model explained task effects. For the different model variations, we present results from likelihood-based model comparisons. A statistical analysis is also applied to generated scan path data and compared to experimental data, which confirm adequacy of the model fits. Finally, we discuss our results with respect to task dependence on scene-viewing and saliency modeling, inter-individual differences, and more general aspects on process-oriented modeling.

\section{SceneWalk: A framework for dynamical scan-path modeling}
\begin{figure}[t]
\caption{\label{fig:sw_1}
Two-stream architecture of visual attention and inhibitory tagging.}
\unitlength1mm
\begin{picture}(150,100)
\put(15,0){\includegraphics[width=0.77\textwidth]{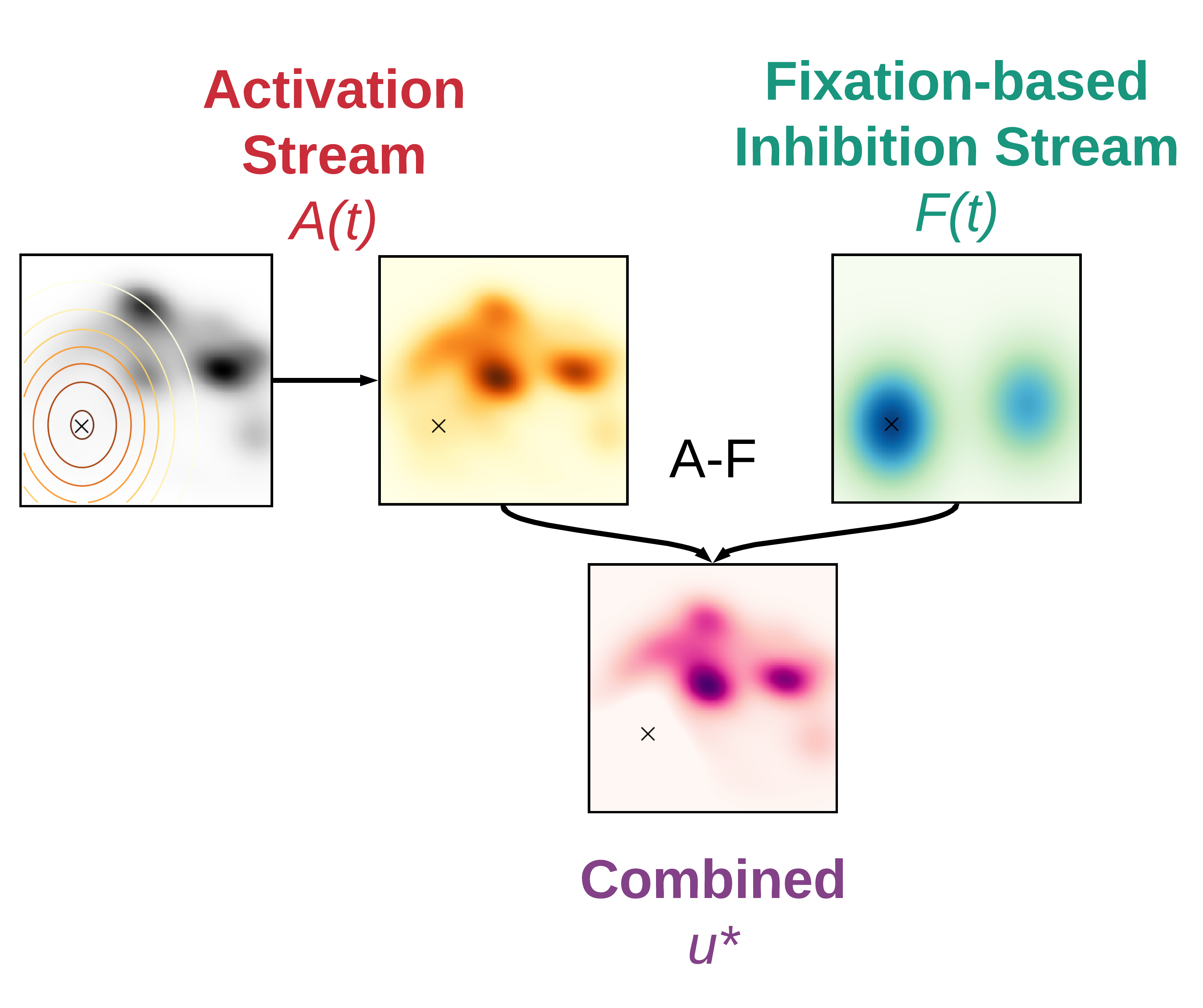}}
\end{picture}
\figurenote{The current fixation position is marked by the symbol ``{\sf x}''. The streams evolve neural activations independently over time depending on the fixation position, input and decay. The activation stream receives as input a saliency map (black and white color map) which is convolved with a Gaussian aperture to approximate the visual attention span (orange color maps). The blue color maps represent inhibitory fixation tagging, which keeps track of previously visited locations. When both maps are combined the result is a priority map we interpret as the fixation selection probability.}
\end{figure}

The SceneWalk model \citep{Engbert2015,Schuett2017} implements two largely independent processing streams: one activatory and one inhibitory (see Figure \ref{fig:sw_1}). Both streams are grounded in theoretical \citep{Itti2001} and experimental work \citep{Rothkegel2016}, showing that they represent the two main factors contributing to fixation selection.

The activation stream combines information about image features with a mechanism for foveation, and thereby yields an approximation of the information that can be extracted from an image at a particular fixation location. Image features include edge and contrast information as well as more high level information such as objects. This information is passed to the SceneWalk model in the form of a normalized fixation probability map. The SceneWalk model is solely a model of dynamics and requires a saliency map to be provided as input for each image. This map could be computed by one of the implemented saliency models \citep[e.g.][]{Kuemmerer2015} that follow the general modeling approach \citep{Itti2001}. For the later interpretation of results, strengths and weaknesses of the saliency models must be separated from shortcomings of the scan path model. Therefore, we will use the experimentally observed fixation density estimate, which represents the theoretical upper limit for the performance of the salience model. Mismatches between data and model output are therefore predominantly caused by the scan path model, although the time-averaging assumption for the fixation density is another approximation that may contribute. The second component of the activation stream is related to the visually attended region. When attention is aligned with the current fixation position, the decreasing receptor density of the retina towards the periphery leads to a decline in visual acuity, which we implement as a Gaussian window centered around the current fixation. For attention shifts to the periphery, discussed below, we keep the Gaussian window approximation to implement an attentional spotlight on an upcoming target \citep{Shulman1979,Engbert2011,Itti2001,Tsotsos1990}. The convolution of saliency and the Gaussian window results in the input to the activation stream \citep{Engbert2015}.

The inhibition stream of the SceneWalk model is responsible for fixation tagging, i.e., keeping track of fixated regions and preventing the continuous return to the same high saliency regions \citep{Klein2000, Bays2012}. Evidence from visual search \citep{Posner1985} and also scene viewing \citep{Klein1999, Bays2012, Rothkegel2016} and electrophysiology \citep{Hopfinger1998, Mirpour2019} shows the relevance of inhibition of return for scan path statistics. The input to the inhibition stream for fixation tagging is also implemented as a Gaussian centered at the current fixation location (see Figure \ref{fig:sw_1}).

The two separate streams evolve continuously over time. Among the previous modeling results, we found that the build-up of activation in the inhibition stream is slower than the activation stream \citep{Engbert2015}. As a consequence, inhibitory tagging evolves slowly, so that refixations of recently fixated scene regions are still possible \citep{Smith2009}. In the activation stream of the most recent version of the model \citep{Schwetlick2020}, we added a directed, smaller facilitation of return \citep{Luke2013, Smith2009} in addition to the slow, global inhibition of return. The interplay of both mechanisms results in a slower decay of activation at the previous location and briefly enables precisely directed return saccades. Thus inhibition of return, attention, and facilitation of return can coexist by separating their temporal dependence. The extended model is described in more detail in the following section. The combination of the activation and inhibitory streams yields a priority map for saccade targeting \citep{Bisley2019}, which the model uses as the 2D fixation probability map for the selection of the upcoming saccade target. In the following, we discuss the dynamical behavior of both activation and inhibition streams as well as its combination to generate a priority map, which depends on fixation history and indicates the time-dependent probability of the target selection process.

\subsection{Activation dynamics of attention and inhibitory fixation tagging}
The most recent version of the SceneWalk model \citep{Schwetlick2020} implements the two-stream architecture discussed above as well as perisaccadic attentional mechanisms, which are related to saccade preparation and execution. As in the original model \citep{Engbert2015,Schuett2017}, the activation and inhibition streams evolve over time and are be combined mathematically to yield a moment-to-moment priority map \citep{Bisley2019}, from which target locations are selected probabilistically. 

The model is implemented on a $128 \times 128$ grid, where $(x,y)$ give the physical coordinates in degrees of visual angle. The inhibition/fixational tagging pathway is defined as a 2-D Gaussian centered around the current fixation position $(f_x,f_y)$. It evolves over the duration of the fixation according to the differential equation 
\begin{eqnarray}
    \frac{{\rm d}F_{ij}(t)}{{\rm d}t} &=& {\omega_F} \left( \frac{G_F(x_i,y_j;x_f,y_f)}{\sum_{kl} G_F(x_k,y_l;x_f,y_f)} - F_{ij}(t)\right) \;,
     \label{eq:diffF}
\end{eqnarray}
where $F$, denotes the fixation-based inhibition stream, $G_F$ is the Gaussian-shaped activation window with standard deviation parameter $\sigma_F$, and $\omega_F$ is the parameter for the speed of decay.

\begin{figure}[t]
  \caption{\label{fig:phases}
  Temporal sequence of the three phases peri-saccadic attention.}
  \unitlength1mm
  \begin{picture}(150,97)
  \put(0,-3){\includegraphics[width=\textwidth]{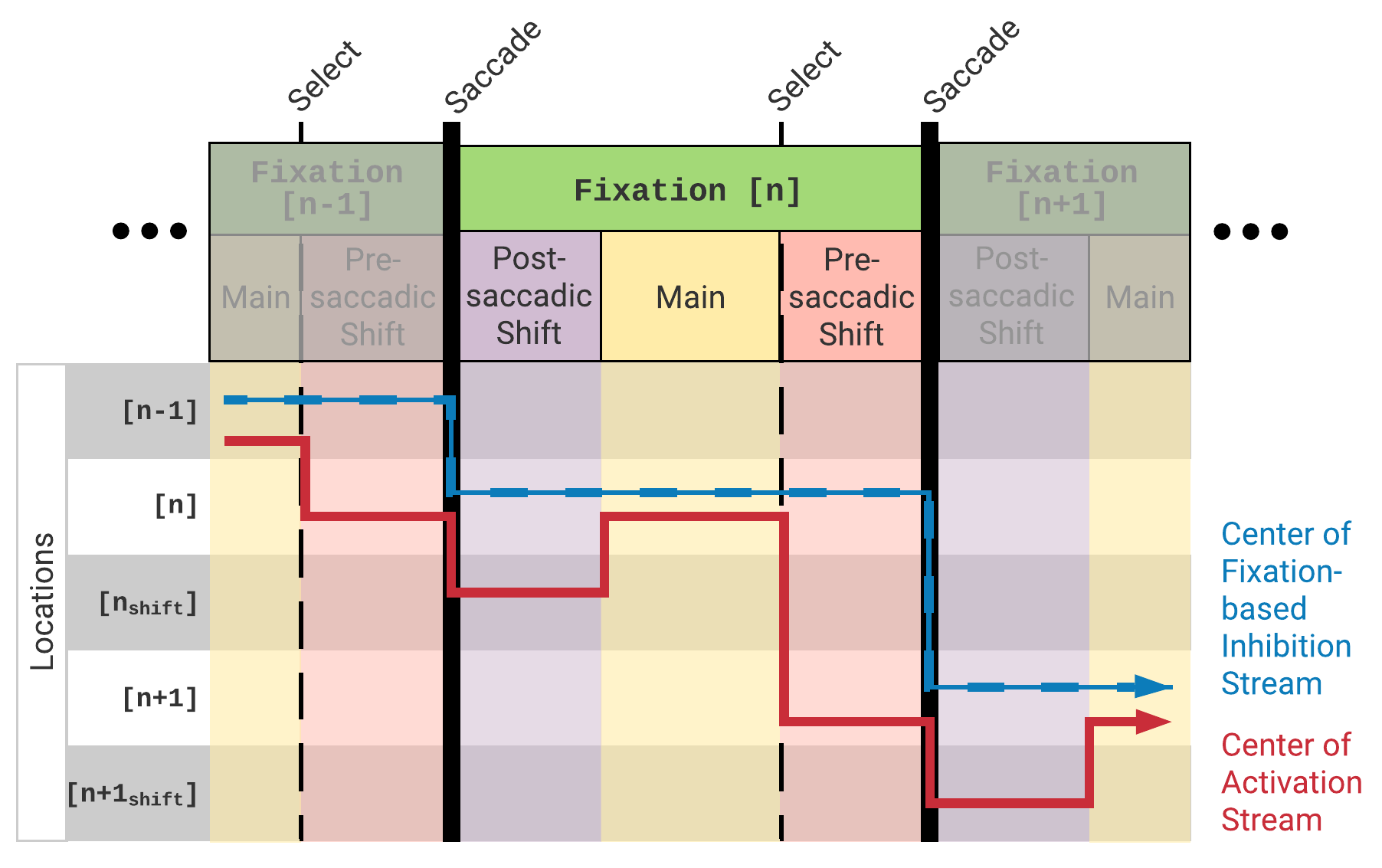}}
  \end{picture}
  \figurenote{In the SceneWalk model, each fixation is split into three phases. During the main phase (yellow color) attention and fixation location are aligned, so that the activatory (red, solid line) and the inhibitory (blue, dashed line) Gaussian inputs are both centered around the current fixation position. The main phase is followed by a pre-saccadic shift (rose color) , where the attention precedes the eye position to the selected location. After each saccade (black line) a brief post-saccadic shift (purple color) causes the attention to be shifted further along the saccade vector, before fixation position and attention once again align at the new fixation position.}
\end{figure}

The activation stream is implemented as a separate ODE, with its own separate time scale. It similarly includes a Gaussian window around the fixation location, emulating the decrease of visual acuity towards the periphery, and includes information about visual saliency, which must be passed to the model. The differential equation for the activation stream is given by
\begin{eqnarray}
    \frac{{\rm d}A_{ij}(t)}{{\rm d}t} &=& \omega_A \left(\frac{S_{ij} \, G_A(x_i,y_j;x_f,y_f) }{\sum_{kl} S_{kl}\, G_A(x_k,y_l;x_f,y_f)} -  A_{ij}(t) \right) \label{eq:diffA} \;,
\end{eqnarray}
where $A$, denotes the activation stream, $G_A$ is the Gaussian-shaped activation window with size $\sigma_A$, centered around the appropriate location for each phase, $S$ is the saliency map of the image, and $\omega_A$ is the parameter for the speed of decay. The computation of numerical solutions of the of ODEs given by Eqs.~(\ref{eq:diffF}-\ref{eq:diffA}) for all grid points $(i,j)$ is discussed in Appendix \ref{App:Model}.

The extended model version \citep{Schwetlick2020} adds changes around the time of saccade to the model, where the temporal aspect is illustrated in Figure \ref{fig:phases}. Each fixation is split into three distinct phases: main phase, pre-saccadic shift, and post-saccadic shift. The rationale behind the extension is that before each saccade, attention precedes the eye to the target location. After a saccade has been executed, research shows evidence for a brief shift to account for the post-saccadic retinotopic attention trace. Thus, in the extended model, the center of the attentional Gaussian does not always align with the fixation position (overt attention), but instead the two decouple around the time of saccade. Previous work has shown that these components of saccade generation improve important statistical properties of the predicted scan path \citep{Schwetlick2020}.

The saccade direction is given by the vector $(x_\delta, y_\delta)$ with $x_\delta=x_n-x_{n-1}$ and $y_\delta=y_n-y_{n-1}$. Therefore, the location of the post-saccadic shift is determined by 
\begin{equation}
    (x_{\rm s}, y_{\rm s}) = (x_{n}, y_{n}) + \frac{(x_\delta, y_\delta)}{\sqrt{x_\delta^2 + y_\delta^2}} \cdot {\eta} \;,
    \label{eq:shiftloc2}
\end{equation}
i.e., the target region of the shift corresponds to a point along the previous saccade vector, where $\eta$ determines the shift amplitude relative to the previous saccade length. 

The peri-saccadic extension of the model requires that the differential equations for the evolution of activations are applied to the three phases for each saccade, since the center of the activation stream is in a different position at each step (see Figure \ref{fig:phases}).

In order to select the next fixation target a priority map is computed, by combining both streams. The exponent $\gamma$ shapes this priority map, making it more deterministic, the higher the exponent becomes, 
\begin{equation}
    \label{eq:subtr}
    u_{ij}(t) = \frac{\left(A_{ij}(t)\right)^\gamma}{\sum_{kl}\left(A_{kl}(t)\right)^\gamma} - C_F \frac{\left(F_{ij}(t)\right)^\gamma}{\sum_{kl}\left(F_{kl}(t)\right)^\gamma} \;.
\end{equation}

Negative values of $u_{ij}$ indicate excess inhibitory activations, which render the saccade targeting probability zero. Thus, for computation of the saccade probability is based on the positive values $u^*_{ij}$, defined as 
\begin{equation}
    u^*_{ij} = \left\{\begin{array}{ll} 
    u_{ij} & \quad\mbox{if}\quad u_{ij}>0 \\
    0 & \mbox{otherwise}
    \end{array}\right.
\end{equation}
Finally, since zero fixation probability does not exist in real experiments, a noise term $\zeta$ is added to warrant fixation in regions with $u_{ij}<0$ with low probability, i.e., the target selection probability at position $(i,j)$ is given by
\begin{equation}
    \label{eq:noise}
    \pi(i,j) = (1-\zeta)\frac{u^*_{ij}}{\sum_{kl}u^*_{kl}} + \zeta\frac{1}{\sum_{kl}1} \;.
\end{equation}

The extended model also includes mechanisms for center bias and facilitation of return, for which we provide detailed mathematical equations in the Appendix.

\subsection{Temporal control of fixation durations and coupling to local saliency}
Because of the dynamical nature of the activation maps in the SceneWalk model, saccadic selection probabilities (or the priority map) change over time during fixations. Therefore, we clearly expect that the model to predict interactions of temporal and spatial aspects of saccade preparation. This theoretical expectation is in good agreement with the results of statistical parameter fitting in the LATEST model \citep{Tatler2017}, which demonstrates various correlations between spatial and temporal aspects of saccade selection. 

We assume that fixation durations are controlled by a continuous-time discrete-state random walk process \citep[see also][]{Laubrock2013,Nuthmann2010}. The distribution of fixation durations $T$ generated by this random walk is a Gamma distribution, which can be written as
\begin{equation}
	g(T) = \frac{\displaystyle b^q}{\displaystyle\Gamma(q)}T^{q-1}e^{-bT},
	\label{eq:gamma}
\end{equation}
with free parameters rate $b$ and shape $q$. The mean fixation duration is given as $\mu_T=q/b$ and its variance is $\sigma^2_T=q/b^2$. It is important to note that our model does not critically depend on the assumption of a Gamma distribution. The broad class of information accumulation or drift-diffusion models \citep{Smith2004} generates qualitatively very similar distributions. The WALD-EM model by \citet{Kucharsky2021} assumes a WALD (inverse Gaussian) distribution, which gives a comparable goodness-of-fit to the experimental data.
Thus, the total duration of each fixation is generated by sampling from the Gamma distribution. According to the most recent version of the SceneWalk model \citep{Schwetlick2020}, a fixation is subdivided into post-saccadic, main, and pre-saccadic phases. The full duration is therefore split into the three phases. The post- and pre-saccadic phases have fixed durations, which is an assumption inspired by experimental work on predictive allocation and remapping of attention \citep{Rolfs2011}.
Specifically, the duration of the shifts were set to $\tau_{pre}=0.05$~s for the pre-saccadic shift and $\tau_{post}=0.1$~s for the post-saccadic shift, corresponding to the approximate durations found in the literature \citep{Rolfs2011, Golomb2008}.

An interesting and important question is if and how local saliency and mean fixation duration are related. Here we assume that mean fixation duration $\bar{T}_i$ at fixation location $x_i$ parametrically depends on the logarithm of the local saliency $\log s(x_i)$. We assume that the shape parameter $q$ of the distribution is constant, while the rate parameter $b$ varies in relation to current input. Therefore, we will try to estimate the parameters $t_\alpha$ and $t_\beta$ for a linear relationship between parameter $b$ and the logarithm of the local saliency, i.e.,
\begin{eqnarray}
\label{eq:parb}
	b = t_{\alpha} + t_{\beta} \log s(x_i) \;.
\end{eqnarray}
In principle, we assume that the model's activation value at location $x_i$ should be used in Eq.~(\ref{eq:parb}), not local saliency. For simplicity, the current version of our model uses the (logarithm of the) local saliency as an approximation for the average local activation. 

In the following sections we will refer this, most recent, version of the model (with timing mechanism and attentional shifts) as SceneWalk. Previous versions of the SceneWalk model are not subject of this paper.

\subsection{Full likelihood function for fixation positions and fixation durations}
Previous versions of the SceneWalk model did not explicitly model saccade timing. With the gamma-distributed random-walk process for saccade triggering, we follow a strategy similar to the LATEST model \citep{Tatler2017}. In this section, we derive the full likelihood function of the model by including fixation durations in the likelihood function. A similar approach has been developed in the WALD-EM model \citep{Kucharsky2021}, who used a combined spatiotemporal likelihood. As discussed below, the spatial and temporal likelihood can be factorized, so that the log-likelihood sums up from spatial and temporal contributions.

A fixation $i$ is determined by position $x_i$ and fixation duration $T_i$, i.e., $f_i=(x_i,y_i,T_i)$. A scan path is a fixation sequence ${\cal F}_N=\{f_1,f_2,...,f_N\}$ of $N$ fixations. For an experimentally observed (or simulated) sequence of $N$ fixations, the log-likelihood $l_M(\theta|\mbox{data})$ under model $M$ specified by parameter vector $\theta$ is given by 
\begin{equation}
\label{eq:basicloglik}
l_M(\theta|\mbox{data}) = \sum_{i=1}^N \log P_M(f_i|{\cal F}_{i-1},\theta)	 \;,
\end{equation}
 where ${\cal F}_{i-1}$ is the fixation sequence up to fixation $i-1$. The probability $P_M(f_i|{\cal F}_{i-1},\theta)$ can be decomposed into a spatial (fixation location $x_i$) and temporal (fixation duration $T_i$) part, i.e.,
 \begin{equation}
 P(f_i|{\cal F}_{i-1},\theta) =  P^{\rm spat}(x_i,y_i|{\cal F}_{i-1},\theta)\cdot P^{\rm temp}(T_i|x_i,y_i,{\cal F}_{i-1},\theta)	\;.
 \end{equation}
Therefore, the log-likelihood can be written as
\begin{equation}
\label{eq:loglikdecomp}
l_M(\theta|\mbox{data}) = \sum_{i=1}^N \left( \log P_M^{\rm spat}(x_i,y_i|{\cal F}_{i-1},\theta) + 
  \log P_M^{\rm temp}(T_i|x_i,y_i,{\cal F}_{i-1},\theta) \right)	\;.
\end{equation}
With the general procedure for sequential likelihood computation given by Eq.~(\ref{eq:sequenlik}), we can write the log-likelihood of a full fixation sequence $F_N$ as 
\begin{equation}
\label{eq:loglikscan}
l_M(\theta|{\cal F}_N) = \sum_{i=2}^N \log P_M(f_i|{\cal F}_{i-1;\theta}) \;.
\end{equation}
Therefore, this spatio-temporal log-likelihood expands on and replaces the original, purely spatial likelihood function described in \citet{Schwetlick2020}.

\subsection{Computational Bayesian inference of the SceneWalk model}
\label{sec:CompBayes}
With the computation of the model's likelihood function described in the previous section, Bayesian parameter inference can be implemented on a computer \citep{Schuett2017}. The advantage of the Bayesian framework is that we estimate not only point estimates for each parameter, but have access to the full posterior distribution over the model parameters. This is particularly desirable in models with a complex likelihood structure, where posteriors may be multi-modal or when we are interested in how much the data constrains the parameters \citep{Schad2021, Gelman2013}. Past studies have yielded promising results when applying Bayesian methods to dynamical cognitive models \citep[e.g.][]{Schuett2017,Rabe2021,Seelig2020,Kucharsky2021}. For example, model parameters could be estimated for single participants, which was impossible before.

The most common numerical method for computation of the posterior is using Markov Chain Monte Carlo (MCMC) sampling \citep{Gilks1995}.  A version of this general approach is a random walk that samples higher density regions of the target distribution more frequently than lower density regions \citep{Brooks2011}. Beginning in a random location, the algorithm selects a candidate point according to a proposal distribution around the current location. This point can then be accepted or rejected based on the likelihood value at that location. It is important to note that even low probability points can be accepted. Thus, the algorithm proportionally samples the target distribution \citep{Brooks2011}. 

In the present study, we applied the differential evolution adaptive Metropolis (DREAM) algorithm \citep{Vrugt2011}, which is a general-purpose MCMC sampler with excellent performance on complex, multimodal problems. The DREAM algorithm runs multiple Markov Chains in parallel, which can exchange information about past states. The latest version MT-DREAM(ZS) combines the strengths of multiple-try sampling, snooker updating, and sampling from an archive of past states \citep{Laloy2012}.  These improvements help to optimize the convergence rate and also reduce the probability of individual chains running out of bounds or getting caught in local maxima. Recently, we applied the DREAM(ZS) algorithm successfully to the previous model version  \citep{Schwetlick2020}.
 
For the purposes of examining the differences between tasks, we split the data into a training and a test set. Thus, for each participant and task a randomized subset of $3/4$ of the trials are considered training data and $1/4$ is considered test data. For the parameter inference, we use training data. The sequential likelihood for each fixation in the training data is calculated for each point in the parameter space sampled by the estimation algorithm.

\begin{table}[t]
\caption{\label{tab:modelparameters}
Model parameters for numerical inference\\[1ex]}
\scalebox{1.0}{
\input{Tab1.txt}
}
\figurenote{Range, mean, and standard deviation (SD) specify the truncated Gaussian priors for each parameter.}
\end{table}

We estimate a subset of all model parameters that turned out to be critical for reproducing the most important statistics in experimental scan paths during previous studies \citep{Schwetlick2020}. An overview of all fitted model parameters is given in Table \ref{tab:modelparameters}. Priors were informed by the previous work with the model on other data sets. We used truncated Gaussian distributions as priors and kept them relatively uninformative in order to allow the data to constrain the model freely for each subject. The prior parameters are also reported in Table \ref{tab:modelparameters}.

\section{Experiment}
With the theoretical extension of the SceneWalk model to generate fixation durations via explicit timing of saccades we set out to investigate a model-based explanation of task effects during scene viewing. The experimental data are taken from a recently published paper that report results from a paradigm with different viewing tasks \citep{Backhaus2019}. The experimental study includes eye-tracking data from 32 participants with normal or corrected-to-normal vision in a scene viewing experiment. Participants were asked to solve four different tasks while viewing 30 natural images. 

Here, we focus on a basic description of  the viewing tasks. For more details about the original experiment see Appendix \ref{App:Exp}. Participants were required to count the number of people in the scene images (Count People). Each image contained between 0 and 9 people; in some cases people were  well hidden in the pictures. Another count task was to determine the number of animals shown in the image (Count Animals). Again, the number could vary between 0 and 9 animals. Since animals can appear in very different shapes and places compared to humans, the authors assumed that counting animals is the more difficult task. Both counting tasks share some characteristics of search tasks \citep{Backhaus2019}, because of the necessary detection of object type before counting. 

The remaining two tasks investigated by \citet{Backhaus2019} are more unspecific with respect to the relevant scene regions, since in these tasks, participants were asked to guess the time of the day an image was taken (Guess Time) and to guess the country where the image was taken (Guess Country). The authors expected that light and illumination, the actions shown in the image (e.g., having lunch) but also clothing could give clues to the time of day or country of origin. These less specific viewing tasks might be looked upon as mildly constrained free-viewing tasks, while the more specific counting tasks might be considered as approximations to search tasks. Across all participants, the four tasks were performed for each image. While each individual participant solved all four tasks, only two of the tasks were solved for the same image in a randomized order.

The experimental data were used to explore how different task instructions influence model parameters. It is important to note that our approach required that time-ordered scan paths for each trial are available, i.e., a sequence of $N$ fixations, ${\cal F}_N=\{f_1,f_2,...,f_N\}$, to evaluate the model. Each fixation $i$ is a combination of fixation location $x_i$ and fixation duration $T_i$. For the scan path ${\cal F}_N$, the log-likelihood is computed using Eq.~(\ref{eq:basicloglik}). In order to limit the variability in scan path lengths we limited the maximum number of fixations per trial to 20, i.e., we removed the last fixations of a trial where necessary.

Experimental data were split into a training and a test sets. For the fixation densities that serve as input saliency maps to the SceneWalk model we used different data for general and task specific densities. The task specific saliency maps were estimated from all fixation sequences obtained from the corresponding task condition. The general saliency maps were computed based on experimental data from a separate study in which the same images were shown in a free-viewing paradigm \citep{Backhaus2022}.

\section{Results}
The key motivation for the current study was a model-based analysis of the influence of task on viewing behavior in natural scenes. Results from statistical model inference may be investigated at three different levels. First, we analyze the parameter values (obtained from the training data), which translate into process assumptions as they possess specific interpretations in our mechanistic model. For example, numerical values must fall within a range that is defined by its interpretation. Second, the model likelihood for the test data set indicates the quality of the fit and will be used to compare model variants. Third, we compare model-generated data to experimental data. In Bayesian analysis, this step is termed posterior predictive checks \citep{Schad2021}. Related analyses are highly indicative of which behavior the model captures well and which aspects might be caused by yet unidentified mechanisms. Capturing interindividual differences will be an important criterion for our model. The workflow for our dynamical modeling study is summarized in Figure \ref{fig:infographic} in Appendix \ref{App:Info}.

\subsection{Parameter estimation}
In order to fit the parameters of the model to the task-dependent scene-viewing study \citep{Backhaus2019}, we implemented a Bayesian workflow as proposed by \citet{Schad2021}, for which the likelihood computation for each scan path, Eq.~(\ref{eq:loglikscan}), is an essential prerequisite. For MCMC sampling we used the PyDREAM implementation \citep{PyDREAM_Shockley2018} of the DREAM(ZS) algorithm \citep{Laloy2012}. Based on priors for model parameters (see Table \ref{tab:modelparameters}) informed by previous studies \citep{Schwetlick2020,Schuett2017}, PyDREAM generates samples converging to the posterior distribution over the model parameter space. We ran 3 chains of 20,000 iterations for 9 parameters for each of 32 participants in each of 4 tasks. These numerical computations were carried out for both the model variant with task-dependent saliency input maps and for the model variant with one general-purpose saliency input for all tasks. Thus, we report data from 256 model fits of 9 parameters each. As suggested by \citet{Vrugt2011} we verified the convergence of the estimation using the Gelman-Rubin $\hat{R}$ statistic (see Appendix \ref{sec:appx_conv}, Figure \ref{fig:conv}). \footnote{Full details of the implementation of the model, the inference, and a variety of checks and to ensure correct behavior are included in the OSF repository.}

Due to a combination of the number of models, model parameters, and number of iterations for each scan path, we conducted parameter estimations on a medium-size multi-core system. The sequential nature of scan paths computations allows parallelization of the iterations between scan paths but not within. One likelihood evaluation, i.e., one iteration in the MCMC sampling algorithm, can be computed within about 10 seconds. One model (out of the total 256), using 28 CPUs, with three parallel chains of 20,000 iterations required an approximate computing time of 55 hours.

\begin{figure}[p]
\caption{\label{fig:complete_fits}
Marginal posteriors for all estimated model parameters across the four tasks}
  \begin{center}
    \includegraphics[width=\textwidth]{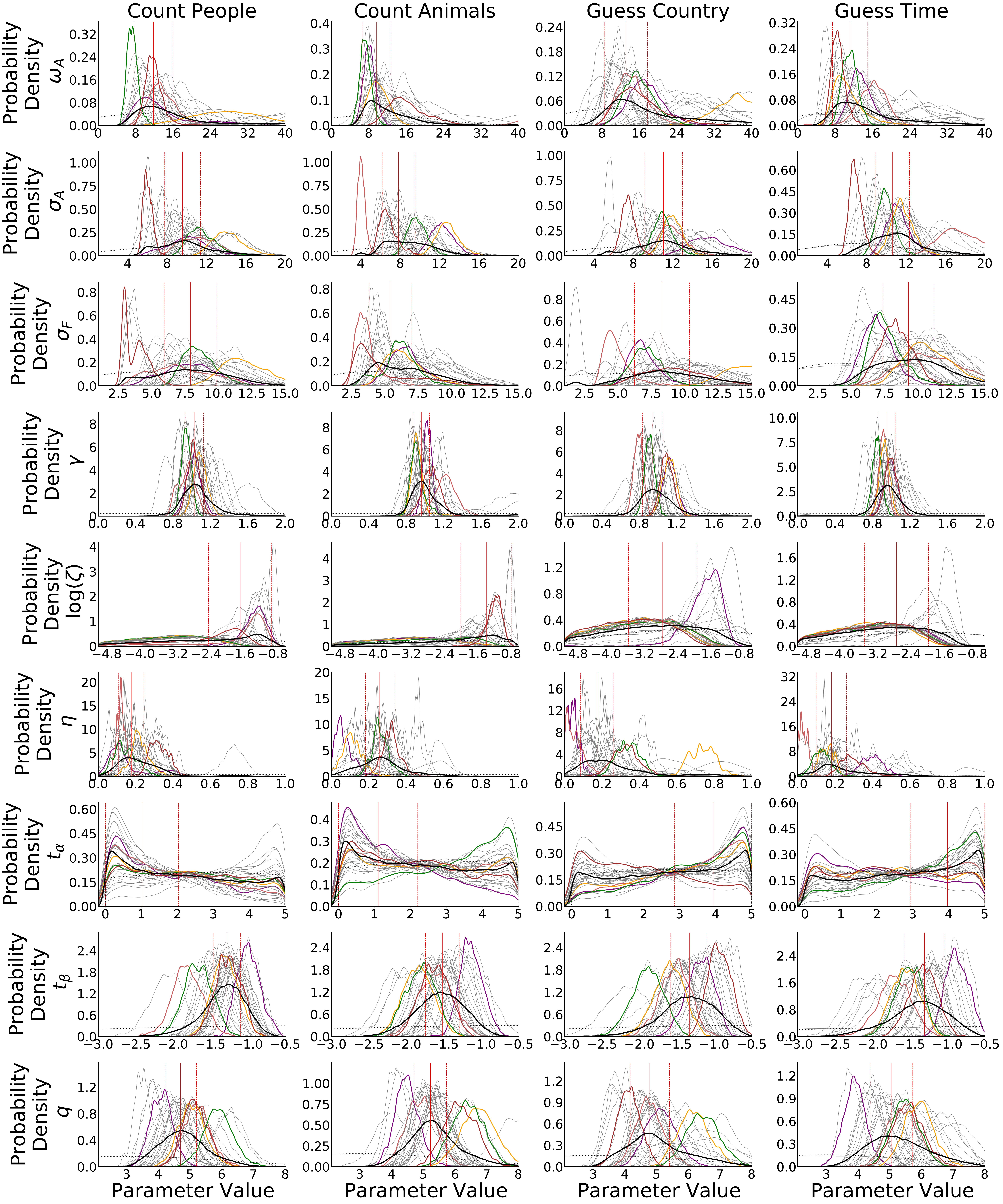}
  \end{center}
\figurenote{The panels visualize the marginal posteriors for all estimated model parameters using task-specific saliency maps. The columns indicate the four tasks; rows represent the 9 estimated model parameters. Each grey line is one subject. The colored lines correspond to data from arbitrarily selected participants, so that results for some participants can be compared across different parameters and tasks. The black lines represent averages over all participants, i.e., a kernel density estimate computed jointly for the samples of all models which shows the trend and spread of parameter values. The dotted lines visualize the prior distributions. Vertical red lines mark the 50\% highest posterior density interval.}
\end{figure}

In the Bayesian approach, the posterior density  contains all information about the model parameters. Figure \ref{fig:complete_fits} shows the marginal posteriors of all estimated model parameters (Tab.~\ref{tab:modelparameters}) for task-specific saliency maps, as this is our baseline model. The parameters in most cases converge to a distinct posterior distribution, which encode individual differences. As an example, $\sigma_A$ and $\sigma_F$ should be noted as parameters where the differences between the participants are explained as differences in attentional span variability. 

\begin{table}[p]

\caption{\label{tab:params}
Point estimates for each parameter by task.}
\begin{center}
\rotatebox{90}{
\scalebox{0.9}{
\input{Tab2.txt}
}
}
\end{center}
\figurenote{The reported point estimates for each parameter and model are the center of the 50\% maximum posterior density interval, averaged over subjects.}
\end{table}


We now discuss important effects of the task on the SceneWalk model parameters using task-specific and general saliency model variants, as reported in Table \ref{tab:params}. 
Note that in Figure \ref{fig:complete_fits} as well as in Table \ref{tab:params} we report general population-level trends, although the models were fitted individually for each subject and task. To give overall interpretation of results, we averaged the marginal posteriors and report maximum posterior density measures of this average. It is important to note that these approximations are not equivalent to parameters fitted generally to the whole population, and disregard correlations between parameters. This measure is used solely descriptively; all further analyses and statistics were conducted using the full marginal posteriors for each model fits.

The parameters $\sigma_A$  and $\sigma_F$ represent the sizes of the Gaussian-shaped inputs for the activation and inhibition streams, respectively. Both are smaller in the Count conditions than in the Guess conditions, indicating a more localized focus in the Count conditions. In fact the Count Animals condition is characterized by the smallest values for both parameters. Locations of animals in the photographs are very diverse, requiring detailed inspection, and making the task the most difficult of the four.

Parameter $\zeta$ is the noise term. It is larger in the Count conditions, indicating that the data is less predictable in those conditions than in the Guess conditions (note that $\zeta$ is plotted as $-\log(\zeta)$ and is therefore negative; smaller values have larger negative values). This could again be interpreted as the result of the more directed viewing behavior in count tasks.

The coupling of local saliency (or empirical fixation density) and mean fixation duration is an important new component of the SceneWalk model. For all of the four task conditions, the coupling parameter $t_\beta$, Eq.~(\ref{eq:parb}) turns out to be negative with zero outside the credibility interval. Thus, for higher saliency at position $x_i$ compared to position $x_j$, i.e., $0<s(x_j)<s(x_i)<1$, we have $\log s(x_j)<\log s(x_i)<0$. Since $t_\beta<0$, the rate parameter $b$ will be larger at position $x_j$ compared to position $x_i$. Finally, since the mean fixation duration $\mu=q/b$, we obtain a longer mean fixation duration at the high-saliency position $x_i$ compared to the low-saliency position $x_j$. Therefore, in our model, image patches of higher saliency will be fixated longer on average. This is in good agreement with the results obtained for the LATEST model, where the decision rate is negatively correlated with saliency \citep{Tatler2017}.

\subsection{Likelihood for general versus task-specific saliency}
The model likelihood informs about the overall adequacy of the model for explaining the experimental data. An important theoretical question is related to the relative performances of the model variants with general and task-specific saliency maps. For example, are task-specific effects primarily due to task-specific saliency maps or can we find task-specific parameters in scan path generating processes? To answer these questions, we fitted two model variants, one model variant where the input saliency map was computed from the experimental fixation density of a free viewing task and another model variant where each fixation sequence was obtained from  task-specific experimental data.

\begin{figure}[t]
\caption{\label{fig:like_comp}
Spatial and temporal likelihoods for model variants}
\unitlength1mm
\begin{picture}(150,75)
    \put(0,5){\includegraphics[width=\textwidth]{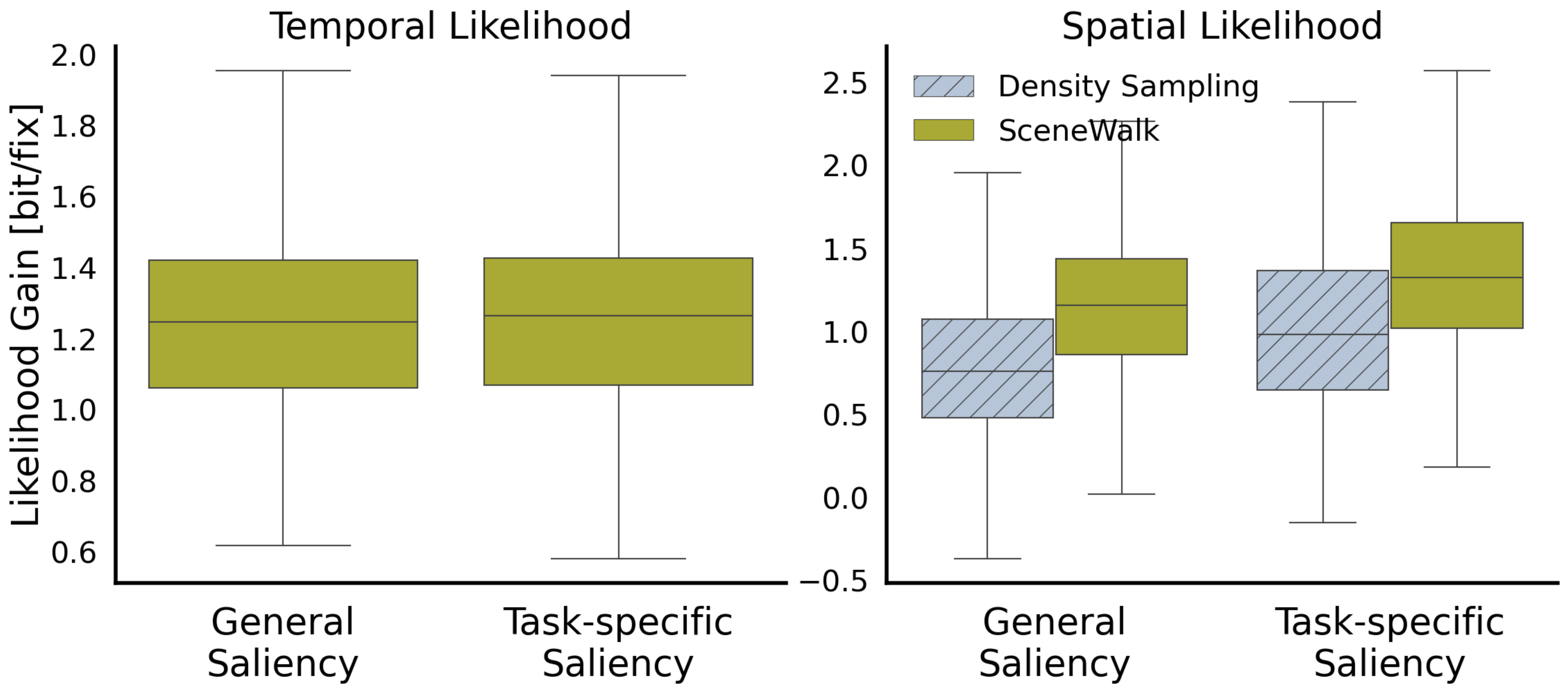}}
    \put(1,7){(a)}
    \put(86,7){(b)}
\end{picture}
\figurenote{Comparison of the model likelihood gain for general and task-specific model variants. (a) The temporal likelihood gain of the SceneWalk model is computed as the difference between the model likelihood and a statistical model (Poisson waiting time distribution). (b) The spatial likelihood gain of the SceneWalk model is obtained as the difference to complete spatial randomness. As a baseline model, we compare the numerical results against a density sampling model (grey, hatched bars) without any dynamics. The combination of the SceneWalk model and the general saliency model outperforms the task-specific density sampling. To improve the visibility of the effects, we omit outlier points in the box plot.The full results of the linear mixed model are supplied in the Appendix (Table \ref{ex:LMM_loglik}). }
\end{figure}

Mathematically, the new model offers an interesting perspective with respect to saccade timing and spatial target selection. We introduced the extended model with an explicit saccade timing mechanism. Based on the model formulation, we showed that the likelihood function can be decomposed into a spatial and a temporal component, Eq.~(\ref{eq:loglikdecomp}). Since saccade timing and spatial target selection should be looked upon as partially independent systems \citep{Findlay1999}, we investigate these two likelihood components separately (see Fig.~\ref{fig:like_comp}).

Figure \ref{fig:like_comp} reports spatial and temporal likelihood components based on the test data. We represent the values as information gain in bit per fixation compared to a random null model. The spatial null model is random selection of points from the grid according to the assumption of complete spatial randomness \citep{Illian2008} with log-likelihood $\log(\frac{1}{128^2})$. The corresponding temporal null model is based on the assumption of a constant probability for saccade onset, which gives a Poissonian waiting time distribution with a rate $\lambda$ corresponding to the average number of fixations per trial found in the data. 

We observe that the models with both task-specific saliency and general saliency can be fitted equally well with respect to the temporal likelihood (Figure \ref{fig:like_comp}a). Even though $t_\beta$ the parameter that couples saliency and durations, is non-zero, the added information of task-specific saliency maps do not improve the temporal likelihood of the model.

In the spatial likelihood, the task-specific saliency maps generate an advantage for the model (Figure \ref{fig:like_comp}b), which can be expected, since the task-specific saliency map is based on specific experimental fixation densities. Sampling from the saliency map is often used to measure for model performance in static models. It is interesting to note that both SceneWalk model versions significantly outperform random sampling from the saliency map. Furthermore, the extended SceneWalk model that uses only general saliency maps outperforms a model that samples randomly from the task-specific saliency map. This result suggests that the dynamical mechanisms in fixation selection are as task-specific as the saliency map.

To analyze the modeling results statistically, we calculated a set of linear mixed models (LMMs) including three fixed-effects \citep{Bates2015}:  Factor Model (Density Sampling vs.~SceneWalk), Factor Saliency (general vs.~task specific saliency), and the interaction of both (i.e., the interaction Model:Saliency). As variance components, we estimated a separate intercept for each subject and for each image. For this analysis we consider values $|t| > 2$ as significant, which produces the following result. Firstly, we find an effect for Factor Model $\beta_{Model} = 0.34~bit/fix$ (using SceneWalk improves the information gain by $0.34~bit/fix$ over Density Sampling). Secondly, we find an effect for Factor Saliency $\beta_{Saliency} = 0.25~bit/fix$ (task-specific saliencies improves the information gain by $0.25~bit/fix$ over general saliencies). Lastly there is no effect for the interaction. Inspection of residuals of the model fit identified 11 outliers out of the total of 1700 data points. We tested a refitted model without the outliers, which did not affect the profile of significant effects. In an additional LMM, we calculated a treatment contrast to statistically validate the difference between 'Task-specific Saliency--Density Sampling' vs.~'General Saliency--SceneWalk'. All comparisons with our baseline 'Task-specific Saliency--Density Sampling' turned out to be significant. The comparison of our main interest revealed a significant difference of $\beta = 0.09 ~bit/fix$ with $t = 2.80$. This significance is compatible with the idea that task dependent scan path dynamics contribute reliably to the model beyond the static task differences (i.e., fixation densities).

\subsection{Posterior predictive checks: Fitting scan path statistics}
\begin{figure}[h]
\caption{\label{fig:systend}
Saccade turning angle as function of saccade amplitude and fixation duration}
\unitlength1mm
\begin{picture}(150,72)
    \put(0,0){\includegraphics[width=70mm]{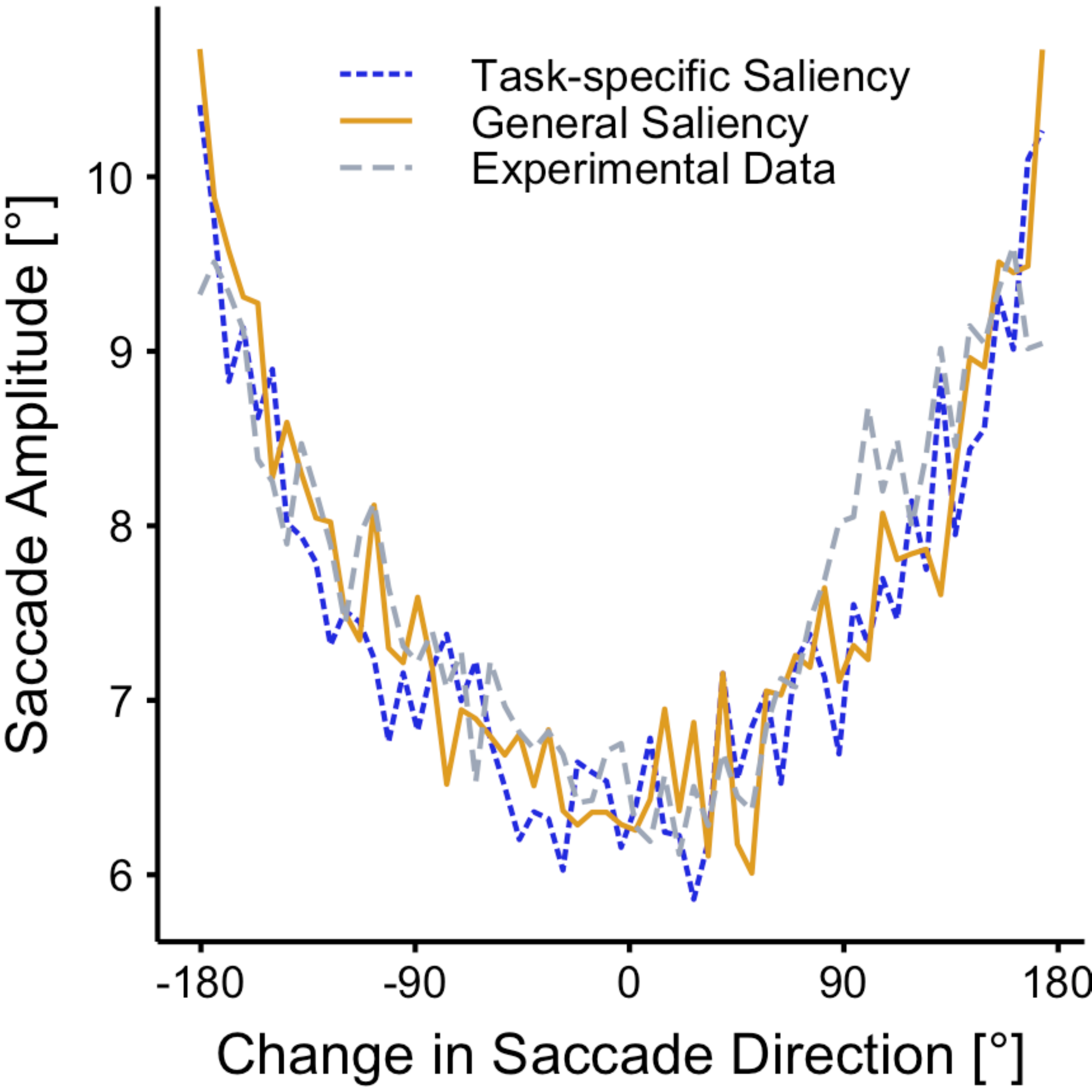}}
    \put(75,0){\includegraphics[width=70mm]{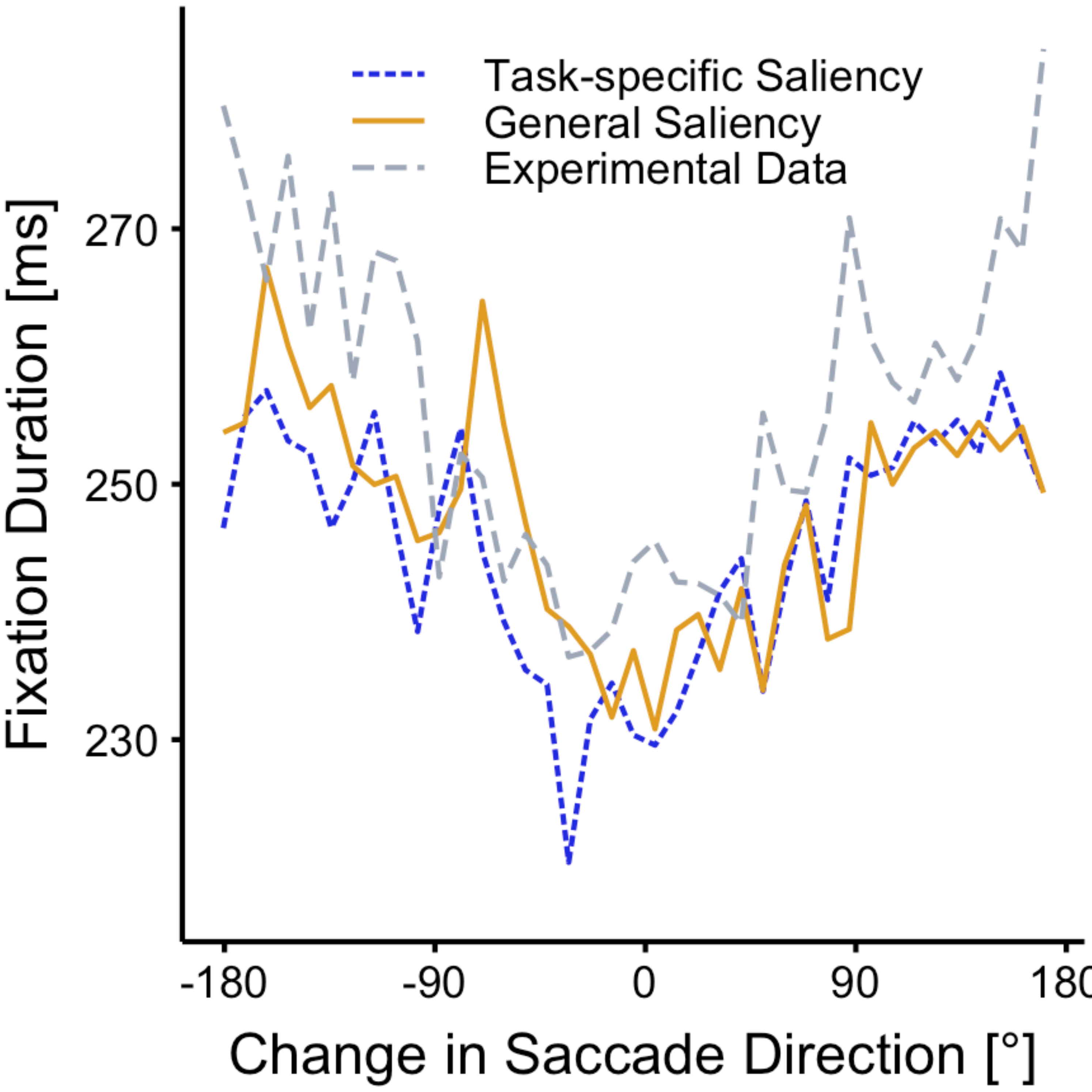}}
    \put(0,-1){(a)}
    \put(75,-1){(b)}
\end{picture}
\vspace{3mm}
\figurenote{(a) Plot of the saccade amplitude as a function of the change in saccade direction (i.e., saccade turning angle). Averages for model variants with general and task-specific saliency (colors) are compared to experimental data (dashed line). (b) Same plot for fixation duration as the dependent measure.}
\end{figure}

Posterior predictive checks refers to the investigation of data generated by the model after parameters have been identified. Model-simulated data may be compared to experimental data in a variety of metrics beyond the model likelihood. As has been shown in previous work \citep{Schwetlick2020}, the SceneWalk model is capable of fitting a variety of metrics of scan path dynamics, beyond mean fixation durations and mean saccade amplitudes as well as their distributions. One important measure of scan path generation is the distribution of turning angles, specifically as a function of saccade amplitude and fixation duration. Here, the posterior predictive checks are important in order to ascertain that our changes to the model architecture did not degrade the fit of scan path statistics with respect to the previous model version \citep{Schwetlick2020}. In Figure \ref{fig:systend} we show that the model fits achieved for this data are well-fitted to the experimental distribution from the test data.

\begin{figure}[t]
\caption{\label{fig:sAcorr}
Correlation between experimental and model-generated saccade statistics}
\unitlength1mm
\begin{picture}(150,73)
    \put(1,0){\includegraphics[width=70mm]{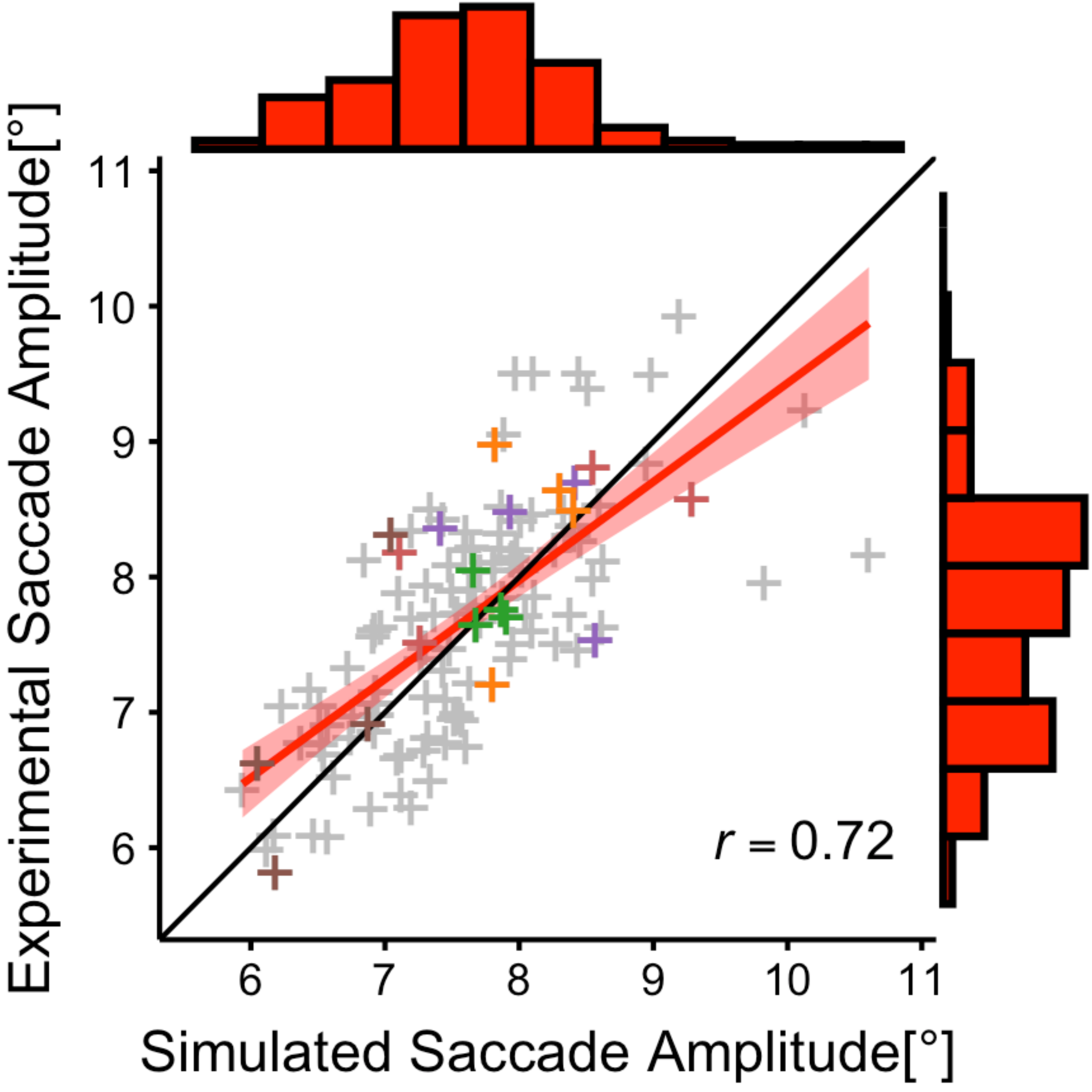}}
    \put(75,0){\includegraphics[width=70mm]{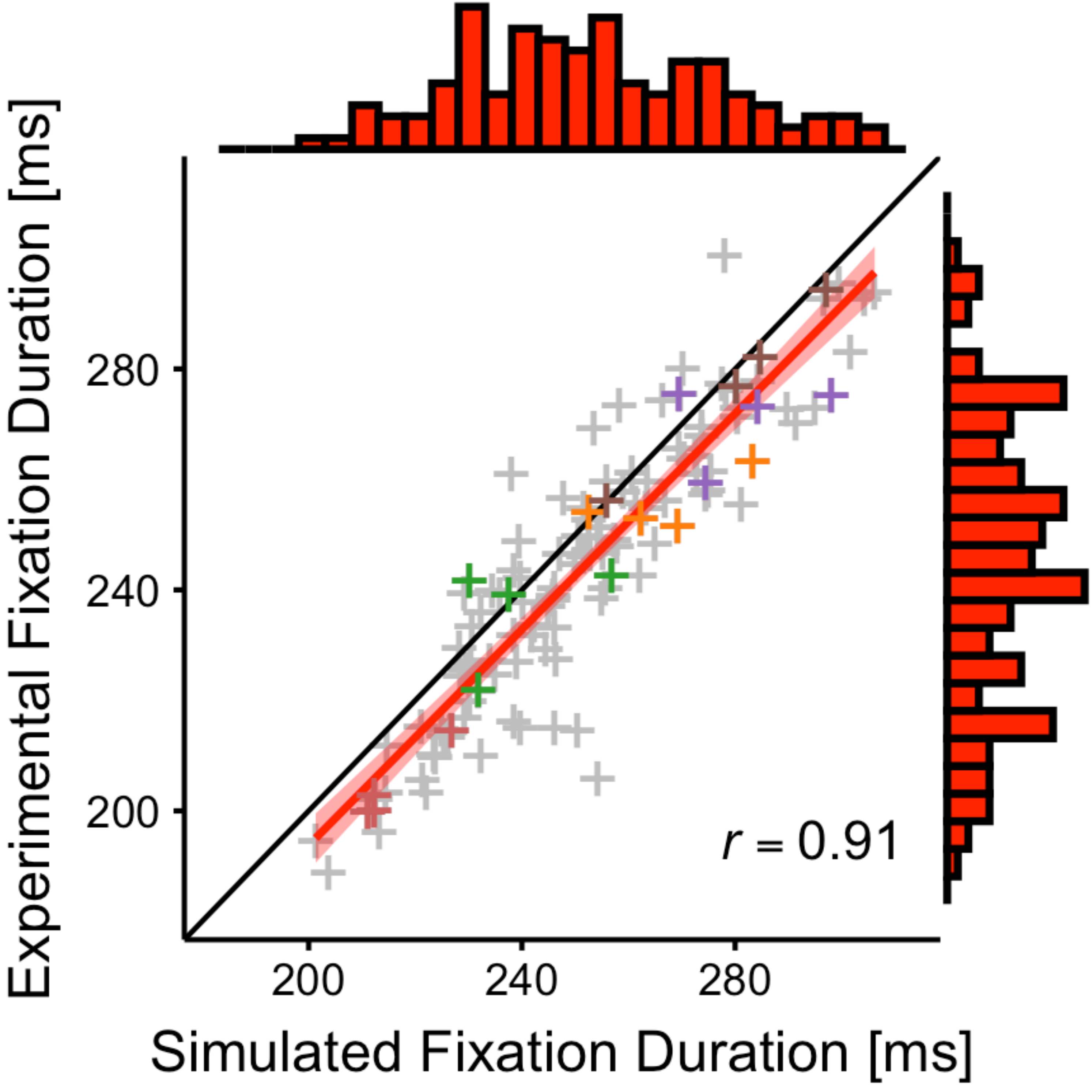}}
    \put(1,-1){(a)}
    \put(75,-1){(b)}
\end{picture}
\vspace{3mm}
\figurenote{(a) Mean saccade amplitude with experimental data plotted along the vertical axis and simulated data plotted along the horizontal axis. Each grey cross ("+") represents experimental and simulated data for an individual observer in a specific task condition. The colored crosses indicate individual observers with the same color mapping as in Figure \ref{fig:complete_fits}. The red line gives the regression line. Histograms at the top and right side of the panel visualize the distributions of saccade amplitudes for simulated and experimental data, respectively. (b) The analogous plot for fixation durations of individual observers in specific task conditions. }
\end{figure}

Generally, posterior predictive checks are necessary for investigating the presence of important characteristics of the experiments in the model-generated data. The above examples for the influences of saccadic turning angle on saccade amplitude and fixation duration are crucial for any scan path model. It is important to note that fitting the model based on the likelihood function and without a consideration of specific ad-hoc metrics produces the correct behavior reliably. As an additional constraint, parameter estimation presented in the current study had access to greatly reduced amount of data compared to previous studies, due to the fact that the model was fit to individual observers on the training subset of a data set with limited trials. We interpret the stable emergence of the critical characteristics of behavior in spite of this as an assurance that the fitting procedure was successful and the experimental data support our model hypotheses strongly.

Figure \ref{fig:sAcorr} shows the correlation of experimental and simulated saccade amplitudes as well as fixation durations for each subject and task, i.e., for each individual set of model parameters. We find a high correlation for both measures, indicating that the model reproduces important summary statistics in the data. Moreover this plot illustrates the way in which the model is able to capture interindividual differences. A model fit to a particular participant who experimentally tends to produce longer than average saccades, will also produce longer saccades when simulating data and vice versa. The same is true for fixation durations. These correlations are an important measure for the sensitivity of the model with respect to interindividual variation. Differences between fits (as shown in Figure \ref{fig:complete_fits}) are not caused by noise or fitting errors but are explaining between-subject variance.

\subsection{Statistical analysis of model parameters from posteriors}
Since in the Bayesian approach we obtain the posterior density over the space of model parameters as a result of model inference, we will be able to run a detailed statistical analysis of the parameter variations across tasks and individual observers. We used linear mixed models (LMMs) to analyze the differences between tasks for each parameter \citep{Bates2015}. As before, we analyzed both models, i.e., the general saliency and the task-specific saliency models.  

\begin{figure}[p]
\caption{\label{fig:nsal_compare} Comparison of parameter estimates between models and the different task conditions.}
\unitlength1mm
\begin{picture}(150,168)
\put(0,0){\includegraphics[width=0.90\linewidth]{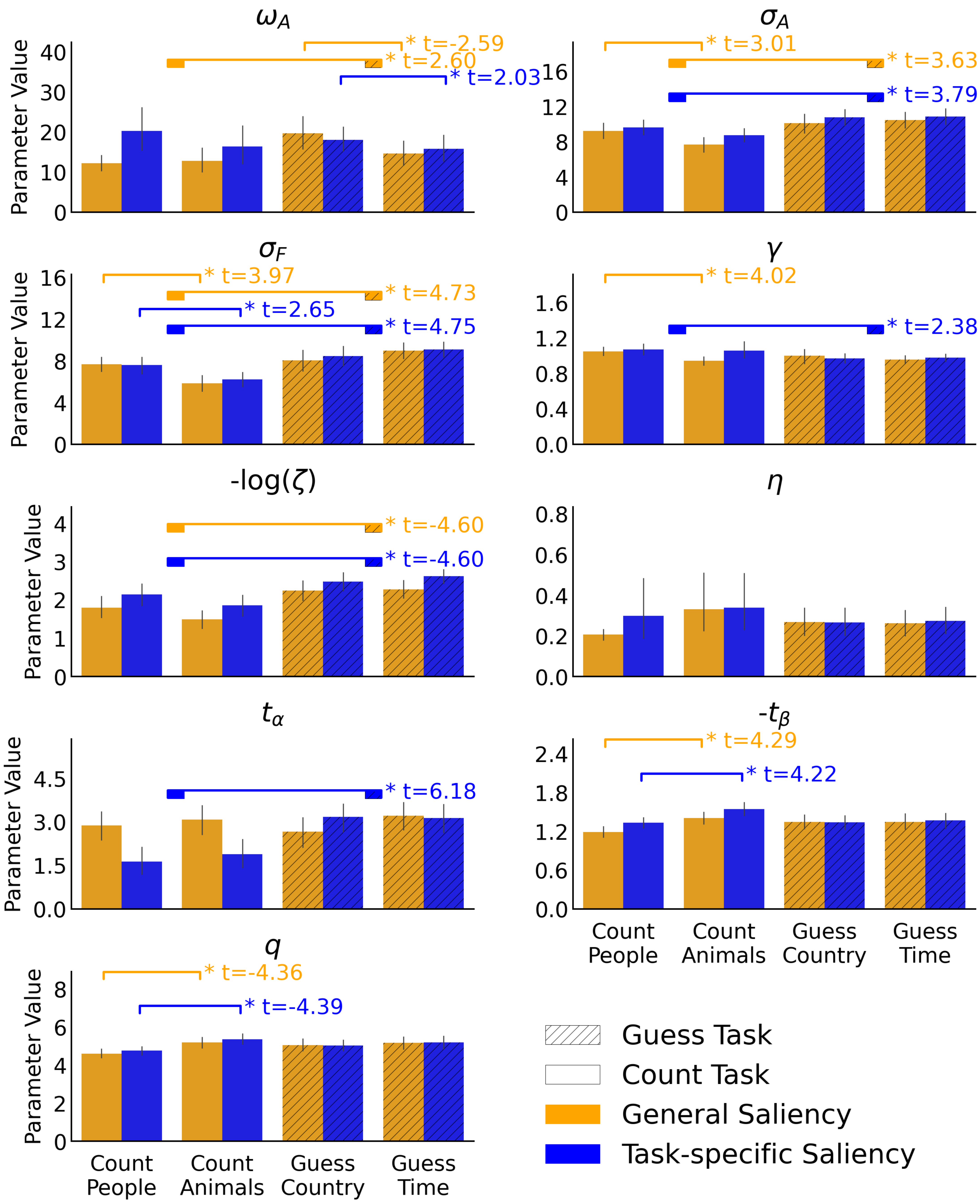}}
\end{picture}
\vspace{4mm}
\figurenote{The orange bars  refer to the general saliency model; blue to the task specific model. The hatching highlights the two Guess tasks. Horizontal lines above the bars show the significant fixed effects as found by a mixed linear model.}
\end{figure}

For the statistical analyses, we sampled parameters from the full posterior density. We ensured the samples were independent by thinning the posterior to every 100th sample and checked statistical independence by analysis of the autocorrelation function. A separate LMM was calculated for each parameter and both the general and specific task models.
The results of these analyses are shown in Figure \ref{fig:nsal_compare}.

The fixed effect structure is taken from \citet{Backhaus2019}, where contrast coding follows the approach of \citet{Schad2020}. We chose a random effect structure with a varying intercept and a varying slope for each contrast by every subject. We did not include image as a factor in the random effect part of the LMM, as we did not model parameters separately for every image. 
The resulting model, presented in the model notation of the lme4 R package \citep{Bates2015,RCore2021}, can be written as
\begin{equation}
\label{eq:LMparameters}\rm
	DV \sim 1 + FGC + FC + FG + ( 1 + FGC + FC + FG || \mbox{subject}) \;,
\end{equation}
where DV represents the dependent variable. The symbol ``1'' represents the model's intercept, FGC denotes the first contrast of both Guess against both Count tasks; FC is the second contrast of Count Animals against Count People conditions; FG denotes the third contrast of Guess Time against Guess Country tasks. The correlations of random effects are not included in the model, which is represented by the double bar sign $||$ in the formula.
 
An important requirement of LMMs is that the residuals are normally distributed. We checked the distributions and calculated an optimal $\lambda$-coefficient via the Box-Cox power transform \citep{Box1964} to re-adjust the experimental data. Even after transform, model residuals of some parameter estimations deviated from a normal distribution. However, \citet{Schielzeth2020} addressed the consequences of violations in distributional assumptions and identified only slightly upwards biases in estimates of varying effect variance. Thus, we expect our results to be reliable in general.


First, referencing Figure \ref{fig:nsal_compare}, we compare fixed effect parameter estimations within the task-specific saliency variation (blue bars). With this model we make the assumption that saliency of image features changes in response to task and ask the question of whether this change in weighting is sufficient to explain the change in behavior. In our analysis, we find differences in parameter values between the two task groups (Guess and Count) for the parameters $\sigma_A$, $\sigma_F$, $\gamma$, $\zeta$ and $t_\alpha$, $t_\beta$, $q$. Above, we qualitatively described the parameters $\sigma_A$, $\sigma_F$, referring to the attentional and inhibitory span, as well as the noise parameter $\zeta$ and timing parameters $t_\alpha$, $t_\beta$, and $q$. In the task-specific saliency model we also find significant differences between Guess and Count tasks for $\gamma$. 

The parameter $\gamma$ controls the weighting the selection map (priority map), making it more or less deterministic. Large values of $\gamma$ lead to steeper peaks in the priority map and thus the target selection is more deterministic. Here, we find that count tasks lead to larger values of $\gamma$ than guess tasks. We relate this finding to the task demands. The object search behavior needed for the Count task, particularly when given a task-specific saliency, is strongly focused on specific targets. The model therefore emphasizes peaks in the selection map, driving more precise and focused target selection by a higher value of the exponent $\gamma$ compared to guess tasks.

Second, we compare within the task groups. As reported in \cite{Backhaus2019}, the two Count conditions themselves evoke different behavior. Searching for animals is a more general tasks (they could be any species, so conceivably found on land or in the air or camouflaged) whereas counting humans is more predictable. Therefore, the difference between these two tasks also caused significant differences in the parameter estimates, specifically for parameters $\sigma_F$, $t_\beta$, and $q$. The model parameter $\sigma_F$, which is responsible for the size of the inhibitory fixation tagging mechanism, is smaller in the Count Animals condition. We interpret this finding by assuming that more local inhibition is particularly important for counting animals to permit finely guided refixations that might be necessary for counting densely packed scene content.

The saccade timing parameters $t_\beta$ and $q$ are also significantly different between the two counting tasks. Specifically, parameter $t_\beta$ determines the influence of the saliency on the duration (the more negative $t_\beta$ the stronger the influence of saliency on duration; see discussion in the section on parameter inference). Beyond the fact that the model parameters $t_{\beta}$ and $p$ reproduce experimental effects on the difference between the two counting tasks, we would also like to point out that saliency maps are more driven by people than by animal locations. When the task involves searching for people, and people cause high saliency, it is most likely to find the search target in high saliency regions. The value of saliency thus influences fixation duration more strongly than in the Count Animals task---a trend that is visible in both fitted parameters $t_{\beta}$ and $p$.

In the next step we investigated the parameter differences when the model was given the same, general saliency map for each task. In this condition too, we find differences between the task groups. Because the saliency itself has smaller explanatory value, the parameters of the SceneWalk model take on a more cogent role. In addition to the significances of the task-specific saliency model described above, we also find significant differences between the task groups in the parameter $\omega_A$. This parameter specifies the speed of the activation decay. Here we find significantly slower decay for Count tasks than for Guess tasks in the absence of task-specific information. We suggest that this may be the case because it is more directly useful in search tasks to keep track of previous locations and significant areas.

The contrast defining the difference between the Count Animals and the Count People conditions is, as in the original analysis by \cite{Backhaus2019}, also significant in some parameters:  $\sigma_F$,  $\gamma$, and $\eta$. Parameter $\sigma_F$, the size of the inhibition Gaussian is smaller for Count Animals condition. This may reflect the size of the objects that are typically being counted. The greater size of $\gamma$ and smaller size of $\eta$ (the length of the post-saccadic shift) in the Count People condition may be related to similar factors of the size and typical locations of the searched objects.

\subsection{Statistical analysis of scan path statistics}
In the previous section, we investigated model-based parameter variations across tasks and observers, which was focus on the models and the meanings of its parameters. Here, we switch to analyzes of the data, where we compare experimental and model-generated data with respect to scan path statistics. 

Based on the estimated parameters per participant, we generated scan paths using the SceneWalk model. These simulated data will be compared to experimental scan paths to investigate whether the statistics of behavior and task differences are reproduced by the model. In the experimental study by \cite{Backhaus2019}, the authors investigated how various scan path statistics, such as fixation duration or saccade amplitude vary with task. Tasks that can be roughly characterized as less-constrained free viewing tasks (here: Guess tasks) produce longer saccade amplitudes and longer fixation durations than tasks with a clear search component (here: Count tasks). 

Previous research has also shown that saccade amplitude and attentional span are related and saccade amplitudes tend to be smaller in search tasks \citep{Trukenbrod2019}. We conducted the same linear mixed model analyses with the original contrast coding and fixed effect structure reported by \cite{Backhaus2019}, but a reduced random effect structure with only varying intercepts for subjects and images (i.e., no varying slopes both on the simulated data and on the experimental data). The resulting model in the model notation of the lme4 R package \citep{Bates2015,RCore2021}, as well as an overview of LMM structures may be found in the Appendix Table \ref{ex:lmm_structure}.

\begin{figure}[t]
\caption{\label{fig:sacstats}
Comparison of fixed-effects from linear mixed model analysis}
\unitlength1mm
\begin{picture}(140,72)
\put(0,0){\includegraphics[width=74mm]{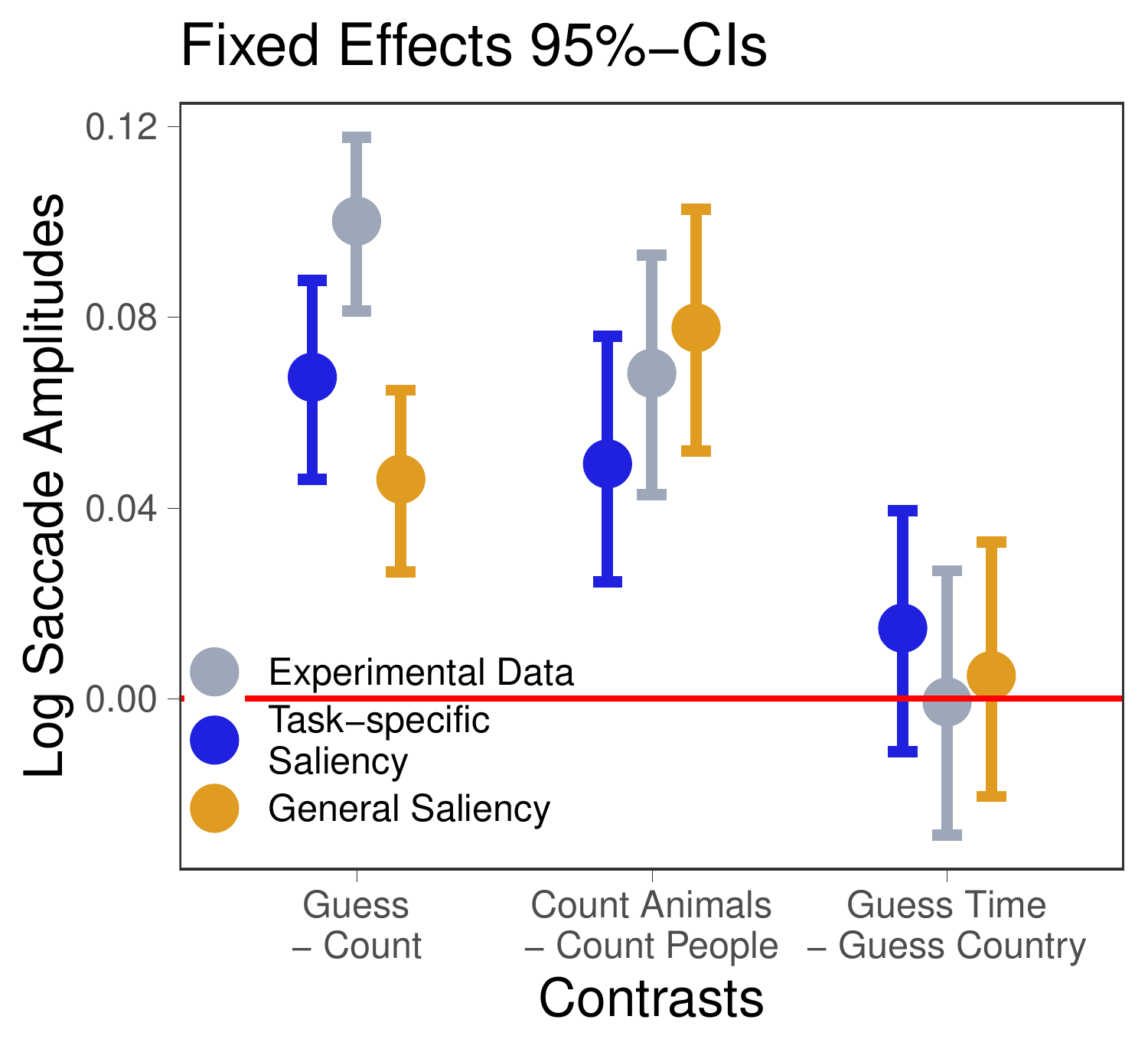}}
\put(77,0){\includegraphics[width=74mm]{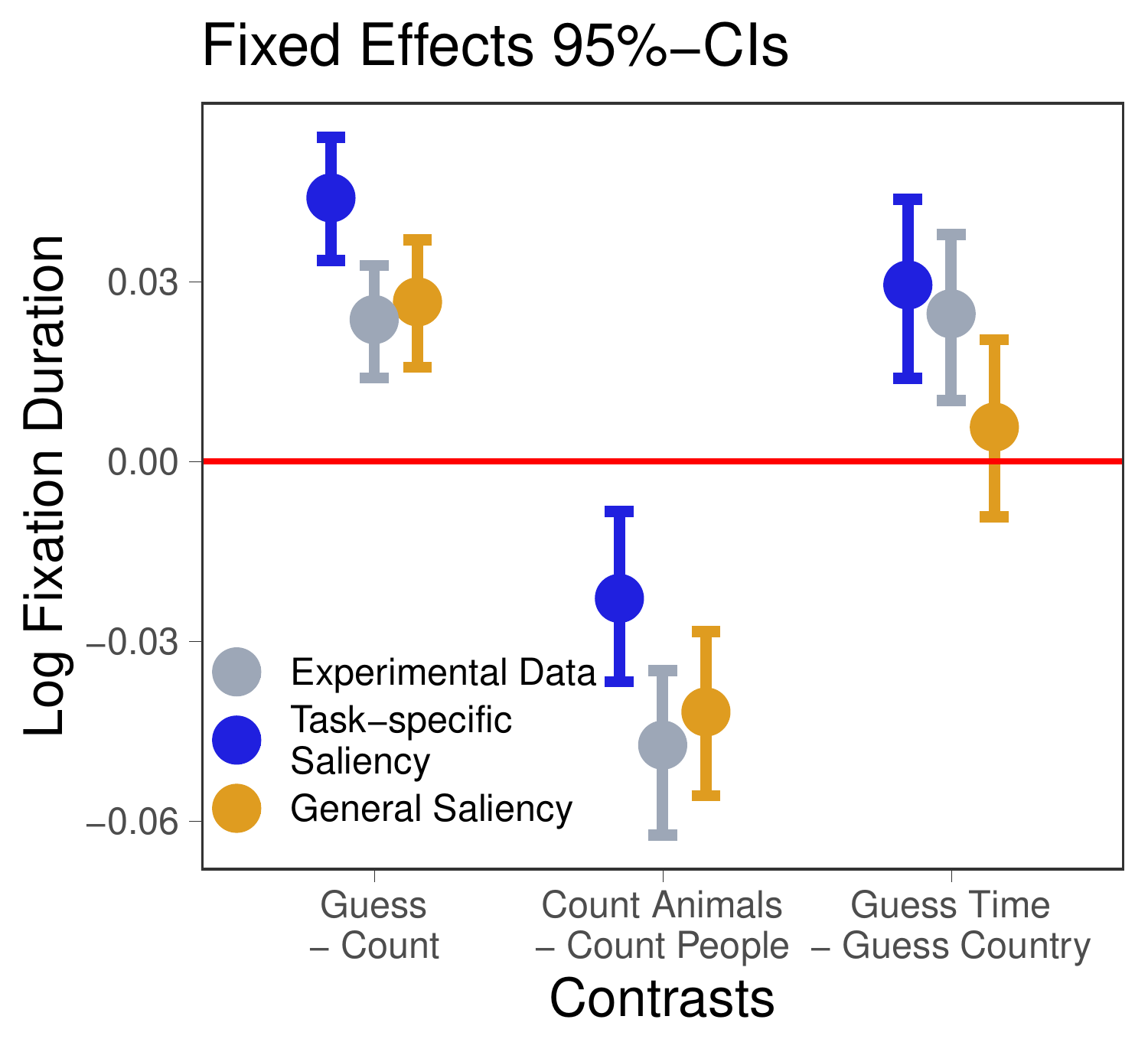}}
\put(0,-1){(a)}
\put(77,-1){(b)}
\end{picture}
\vspace{3mm}
\figurenote{(a) Estimated fixed effects comparison for saccade amplitude analyses. (b) Estimated fixed effects comparison for fixation duration analyses. In both panels, grey color shows the LMM estimates for experimental data, orange color shows the general saliency model, and blue color represents the task-specific saliency model. The horizontal red line marks the zero value, at which there are no differences in the specified contrasts. Confidence intervals around the estimated effects are the bootstrapped shortest 95\% intervals.}
\end{figure}

We found that almost the same contrasts turned out to be significant and the estimated values are in good agreement in all cases (Figure \ref{fig:sacstats}). Results for saccade amplitudes are reported in  Table \ref{ex:LMMSacAmpl}; for fixation durations please consult Table \ref{ex:LMMFixDur}. The only exception is between general saliency, task-specific saliency, and experimental data for the contrast that captures the difference between the two Guess tasks for fixation duration. Specifically, this contrast is significantly positive for experimental and task-specific saliency data. The estimate for the simulated data with a general saliency map, however, can vary to values below zero. Note that the model's responses are slightly muted as compared to the human scan paths.

\section{Discussion}
\label{sec:discussion}
Visual exploration of natural scenes depends on the given objective. This has been noted since the beginnings of vision science \citep{Yarbus1967}. Based on the advances in modeling of visual attention \citep{Itti2001} and eye-movement control \citep{Tatler2017,Schwetlick2020}, we investigated the performance of a computational model of scan path generation for an experiment in scene viewing over four different tasks. We extended the SceneWalk model \citep{Schwetlick2020,Engbert2015} by an explicit saccade timing mechanism and implemented a fully Bayesian framework for dynamical, process-oriented modeling \citep{Schuett2017}. Specifically, in this approach, it is possible to estimate model parameters for individual human observers. Thus, in posterior predictive checks, we were able to carry out a statistical analyses of individual differences across tasks. As a result, we found evidence for specific adaptations of model parameters to task constraints. The extended SceneWalk model reproduces task-effects, individual differences across tasks, and demonstrates an overall advantage for model variants with task-specific saliency maps. 

Overall, our findings suggest that parameters in the generation of scan paths are as highly adaptive to task requirements as are saliency maps. First, given a specific task, human observers seem to adjust the control of saccade dynamics. This is psychologically plausible, since, for example, stronger inhibition of return and smaller saccade amplitudes might contribute to an effective strategy for fine-grained search behavior compared to a less-constrained free viewing task. Second, it is also psychologically plausible that the saliency of certain object features in scene changes with the requirements of the task. Looking for a specific object may result in a strategy of ignoring all features that are unlikely to be associated with that object.  

\subsection{Dynamical modeling of eye-movement control} 
Our current results are an example for process-oriented, dynamical modeling as tool not only for predicting human behavior \citep{Engbert2022}, but also for identifying gaps in our understanding. Over the last decade, major advances were related to model of visual attention in scene viewing, with the  time-independent 2D fixation density as the modeling target \citep{Koch1985, Itti2001, Pan2016, Kuemmerer2015}. Recently, the interest is growing in predicting time-dependent series of fixations, both in the field of vision science \citep{Engbert2005, LeMeur2015, Tatler2017} and in the context of deep learning \citep{Kerkouri2021, Kuemmerer2021}. In our process-oriented approach the SceneWalk model implements specific mechanisms inspired by successful experimental research such as inhibition of return \citep{Klein1999, Klein2000}.  Interestingly, in our model inhibitory tagging is modulated by task, with a smaller spatial size of the inhibitory tagging parameter for counting compared to guess tasks---a finding that underlines the flexibility of the contributing attentional processes in eye-movement control.

An important advantag of the process-based approach over more data-inspired models \citep{LeMeur2015} or deep learning neural networks \citep{Kuemmerer2014} is that there are fewer model parameters in process-based models, which have a clear interpretation with respect to their function in the control of eye movements. Thus, process-oriented models provide insights into how well our current understanding describes the process. Posterior predictive checks, i.e., the comparison of simulated and experimental data along a variety of metrics reveals the gaps between what is implemented in the model and the underlying process. The addition of the new timing mechanism in this work is an example of applying this approach. It is inspired by assumptions from the literature \citep{Tatler2017}, is confirmed by the estimated parameters (the best fit value for the coupling parameter $t_\beta$ is non-zero, indicating that spatial and temporal components are linked), and is validated by posterior predictive checks. Finally, we applied our model and our framework for parameter inference to the estimation inter-subject variability and inter-task variance in scan paths.

\subsection{Model adaptivity: task-specific model parameters}
The SceneWalk model produces specific, systematically different parameter estimations when fit on data from a range of tasks. Using considerable computational resources we conducted separate model fitting procedures for each subject and each task for two model versions. The  parameter estimation successfully found an informative posterior distribution in the majority of cases. These marginal parameter posteriors reveal pronounced differences in value for the different tasks and subjects. The success of the estimation is worth noting  particularly because the amount of data available for the number of estimations was comparatively small. This work contributes insights into the relevance of task and interindividual differences for the process of attention selection. In the next paragraphs we will discuss the parameter differences in detail.

The two most straight-forwardly interpretable parameters in the model are attention span $\sigma_A$ and the inhibition size $\sigma_F$. The estimated parameter values for both are larger in Guess tasks than in Count tasks. We propose to interpret this in the following way. A reduced attentional span enables a detailed inspection of small areas. This is consistent with the finding that search tasks elicit more and shorter saccades. The length of the saccades and the estimated attentional span in our model are highly correlated. For free viewing tasks, a broader attentional span is useful as it allows the viewer to take a wider perspective and take into account more features, but with less detail. In fact we find that the smallest attentional span is found in the count animals condition. This is also the most detail-oriented and difficult task. The inhibition size is also smaller for count conditions. We propose that this is partially a direct result of the amount of inhibition needed to counteract the activation and partially due to a more precise tagging of already-viewed locations. Thus, the parameters reflect the influence of task on spatial gaze statistics \citep{Mills2011, Backhaus2019}.

The parameter $\omega_A$ regulates the speed of the activation decay. The speed of the two streams, activation and inhibition, is separated by an order of magnitude ($\omega_A/\omega_F = 10$). We find a slight difference between Count and Guess tasks for the parameter $\omega_A$ when we fit with one general saliency. Specifically, Count tasks have a systematically slower decay than Guess tasks. We would like to put forward the interpretation that in the case of search tasks the past positions retain more importance. The searcher needs to keep track of already fixated or found objects as well as inhibit discovered distractors. However, we find this effect only when the input saliency is general; no such difference emerges when the saliency map is more informative and task-specific. A possible explanation is that the information which is available longer due to slower decay is related to the information which, in the other case, is present in task-specific saliency maps. That is, the information can either be encoded in the saliency input map or can be accounted for by slower decay( $\omega_A$). In one case the model has to build that representation itself (general case) and in the other it does not need to as the information is in the input saliency (specific case).

\subsection{Temporal control of saccades}
In this study, we we provided an important extension of the SceneWalk model to temporal control of saccades. It is important to note the earlier version of the model \citep{Schwetlick2020} included saccade-related modulations of fixation duration, but not an explicit timing mechanism. The explicit saccade timing enables the model to make predictions not just for the spatial selection of fixation locations and the interaction with fixations, but also for modeling task-dependent, strategic effects in mean fixation durations. 

The new timing mechanism introduces additional variability in the coupling between spatial and temporal selection. The control of fixation durations in scene-viewing were studied earlier based on explicit timing mechanisms \citep{Nuthmann2010,Laubrock2013}. Most recently, the LATEST model combined temporal with spatial aspects of saccade generation \citep{Tatler2017}. While the dynamical part of the LATEST model is limited to the saccade timing, it motivated the integration of a timing component to our fully dynamical framework \citep{Engbert2015,Schwetlick2020}. We successfully implemented a coupling of the local saliency at the current fixation location to mean and variance of the saccade timer. The prior for the spatiotemporal coupling parameter $t_{\beta}$ included the option for this magnitude to be zero, i.e., to infer that saliency has no influence on duration, effectively decoupling the two components.  In accordance with our expectations and with the results of the LATEST model \citep{Tatler2017}, the credibility interval $t_{\beta}$ did not include zero (numerically, the mean is between 1 and 1.5). Thus, we obtained clear evidence for longer average fixation durations at image patches with higher saliency compared to region of lower saliency. 

The likelihood function plays an important role for combined modeling of fixation durations and fixation locations \citep[e.g.,][]{Schuett2017,Engbert2022}. To our knowledge, the first study using spatiotemporal likelihood inference in scene viewing was published by \citet{Kucharsky2021}, in line with conceptual work for eye movements in reading by \citet{Seelig2020}. \citet{Kucharsky2021}'s WALD-EM model combines a standard information accumulation process for saccade timing with a spatial component. Similar to our results, WALD-EM was demonstrated to successfully reproduce several key aspects of eye-movements statistics including interindividual differences. Different from WALD-EM, our model includes biologically motivated, perisaccadic attentional processes around the time of saccade to reproduce several experimentally observed qualitative phenomena such as couplings of fixation durations and turning angles \citep{Schwetlick2020}. Thus, our approach implements a more complicated internal model structure. Both \citet{Kucharsky2021}'s WALD-EM and our SceneWalk model demonstrate the superiority of parameter inference based on a spatiotemporal likelihood function.

\subsection{Interindividual differences in viewing behavior}
An important step forward in dynamical modeling of individual viewing behavior was achieved by the likelihood-based framework for parameter inference. Experimentally, it is well known that saccade statistics and visual attention show marked interindividual differences \citep{Kliegl2010,Makowski2020}. In the past, modeling of an individual observer's behavior was out of reach, since model fitting based on ad-hoc statistics required an amount of data that was typically not provided by experimental studies. As a consequence, model parameters were estimated for data pooled over all of the participants of an experimental study, which precluded modeling of interindividual differences.

As parameter fitting algorithms have improved, it has become possible to reduce the amount of data needed. With the likelihood function available for the SceneWalk model, parameters could be inferred from experimental data on a single-subject level \citep{Schwetlick2020}. Using the task specific data sets in this study, we had to further reduce the amount of data available to our fitting procedure. Fortunately, our MCMC implementation based on the DREAM(ZS) algorithm \citep{Laloy2012}  produced stable posteriors for each individual observer and across tasks.

\subsection{General vs.~task-specific saliency maps}
As in our previous studies \citep{Engbert2015,Schwetlick2020,Schuett2017}, we focus on the investigation of dynamical principles of scan path generation. Therefore, we used experimental density maps as an upper bound for visual saliency models. Because of the available amount of observations, fixation-density maps could be produced from experimental data with specific task instructions.  

One view of task differences in eye-movement control is that the differences mainly occur due to a saliency weighting of different aspects of the image. We might expect, therefore, when using task-specific saliency maps, our model of saccade dynamics to produce very similar parameter estimates for all tasks, since most of the variance in experimental data will be accounted for by the saliency maps. Interestingly, our analyses indicated that saccade dynamics strongly contribute to the adaptive behavior in response to task requirements. Model parameters, e.g., attentional and inhibitory span (i.e., the sizes of the activation and inhibition Gaussians) or parameters related to temporal control of fixation durations turned out to be clearly different between the investigated tasks. 

In addition to this task-specific saliency account, we also identified model parameters in a model variant where the visual saliency was derived from free viewing for all tasks, which we called the general saliency approach. Here, the underlying assumption is saliency is predominantly image dependent and does not change with task. The strongest version of this assumption implies that the observed variation in eye movement behavior is caused by the adjustment of the eye dynamics to task constraints. We found that the model of saccade dynamics still produces reasonable parameter estimates, however, the overall performance of the model was clearly weaker than for the model with task-specific saliency. One might argue that the psychologically plausible assumption would be that adaptation occurs in the saliency map as well as in the eye-movement dynamics. Nevertheless, it still seems very interesting that the model with general saliency outperforms density sampling from task-specific saliency maps. Thus, dynamics contribute significantly to task adaptation. A practical implication of this finding is that in a situation where only general saliency maps are available, adaptation of the eye-movement control system can significantly improve model predictions.

Modern models of visual saliency are usually evaluated with respect to scan paths generated by human observers, and, therefore, will contain both early saliency effect \citep{Itti2001} and high-level influences from scene semantics \citep{Henderson2018}. The same is true of the experimental fixation densities which are used in the SceneWalk simulations. Thus, the question of whether visual saliency is task dependent, is contingent upon the operational definition of saliency.

\subsection{Model performance: posterior predictive checks}
One of the key improvements presented in the study is the likelihood-based parameter inference for modeling individual viewing behavior \citep{Engbert2022}. While likelihood is mathematically rigorous, a maximum-likelihood model's performance can still be poor with respect to qualitative effects. Therefore, we carried out extensive posterior predictive checks, which demonstrated that our model reproduced many of the scan path statistics on the level of individual observers. Moreover, the model explained systematic differences of scan path statistics between the tasks found the in the underlying experimental study \citep{Backhaus2019}. As a dynamical and generative model, the SceneWalk model is capable of simulating scan paths given the estimated parameters and the saliency of an image. We simulated data for each observer and task as well as for both model versions based on general or task-dependent saliencies. Simulated data were compared to the experimental test data. The good agreement between the scan path statistics of simulated and experimental data is an essential component of a psychologically and biologically plausible model. Similarities and dissimilarities allow conclusions about which components of the process are well-captured by the model architecture and which still require explanation. 
In the current study, we report good agreement between model-generated and experimental data. First, we confirm that the same general scan path statistics can be captured well with the extended model that includes temporal control of fixation duration compared to the latest version before this extension \citep{Schwetlick2020}. Second, we compare whether the model captures the task differences found in the experimental data set \citep{Backhaus2019}. The experimental data indicated pronounced differences between the tasks in fixation duration and saccade statistics (e.g., amplitudes). Therefore, our results lend theory-based support to the idea that different viewing strategies are driven by saliency weighting, but also by dynamics of eye guidance.

Further analyses are required to test the model against additional experimental data sets covering a broader variation of task type. In the current work, the model's performance could be improved in view of the comparison to neural network models such as DeepGaze \citep{Kuemmerer2014}. While our model can be fitted to data from individual observers, interindividual variability will be overestimated due to differences in the convergence and identifiability of model parameters. Regularization by hierarchical modeling might be a solution here. Therefore, introducing hierarchical Bayesian dynamical modeling might be another big step forward for modeling individual observer's viewing behavior.

\subsection{Evaluation of our preregistered hypotheses}
Prior to conducting the current study, we preregistered our research plan including the main hypotheses \citep{Schwetlick_prereg}. The first two hypotheses in the preregistration concerned the task influence on attentional span and inhibitory fixation tagging. Specifically, we assumed that the attentional span would be larger in the Guess conditions, which we characterized to be similar to free viewing tasks. Previous research shows that saccade amplitude and attentional span are related and that saccade amplitudes tend to be smaller in Count tasks \citep{Trukenbrod2019}. The results from the estimation of parameter $\sigma_A$ finds support for this hypothesis. Second, we proposed that inhibitory fixation tagging would be more important in search tasks. In fact we find that in count conditions the inhibitory tagging is more directed and less global, i.e. that the parameter $\sigma_F$ is smaller, resulting in more specific inhibitory tagging of regions.

The third hypothesis concerns the decay of past states in the model. We expected for the Count conditions that the decay would be slower compared to the other tasks, since it might be more important to keep track of visited items. In accordance with this idea, we find slightly faster decay in Guess tasks than in Count tasks, as specified by $\omega_A$ in the general saliency model. In the appendix, we provide some more detailed summary and evaluation of our predictions and findings.

\section{Conclusions}
In this work we proposed an advanced model of eye-movement control with application to task-dependent viewing behavior. First, we extended a previous model to include temporal control of fixation durations and the interaction with spatial selection. Second, we applied rigorous statistical parameter inference that showed markedly different results across four different viewing tasks. These findings were corroborated by posterior predictive checks which indicated that these differences also manifest in data simulated by the model fits. Specifically, the model-simulated data reproduced the key scan path statistics found in experimental data. Thus, parameter inference yielded individual parameter estimates not only for tasks but also for each participant in the experimental data. We conclude that the SceneWalk model explained individual differences and task influences on behavior in a theoretically coherent framework.

\section{Acknowledgements}
This work was supported by a grant from Deutsche Forschungsgemeinschaft (DFG), Germany (SFB 1294, project no. 318763901). D.B.~and R.E.~received additional support via grant (EN 471/16-1, DFG).

\printbibliography
\begin{appendix}
\setcounter{table}{0}
\renewcommand{\thetable}{A\arabic{table}} 
\setcounter{figure}{0}
\renewcommand{\thefigure}{A\arabic{figure}} 
\setcounter{equation}{0}
\renewcommand{\theequation}{A\arabic{equation}} 

\newpage
\section{Experimental details}
\label{App:Exp}
\addcontentsline{toc}{section}{Appendix A: Experiment details}
\subsection{Methods}
The eye tracking setup included a mobile eye tracker in a lab with a wide projector screen. Subjects received credit points or a monetary compensation of 10,00\euro\ for their participation. To increase compliance with the task, we offered participants an additional incentive of up to 3,00\euro\ for correctly answering questions after each image (a total of 60 questions). The experiment was carried out in accordance with the Declaration of Helsinki. Informed consent was obtained for experimentation from all participants. The experiment data originally published by \citet{Backhaus2019} are freely available via OpenScienceFramework (OSF, \href{https://osf.io/gxwfk/}{https://osf.io/gxwfk/}).

\subsection{Data preprocessing}
In our laboratory, we developed a processing workflow for the preprocessing of mobile eye-tracking data. Eye movement recordings from our mobile eye tracker are provided in head-centered coordinates. We presented 12 different QR codes around the stimulus material during the experiment. In the video output from the mobile eye tracker, we detected the QR codes using the Pupil Labs software Pupil Player version 1.7.42 \citep{Kassner2014}. The stimulus area within the QR codes is defined as a rectangle. Using a projective transformation provided by the image processing toolbox from MATLAB (The MathWorks, Natick/MA), we converted data points from head-centered coordinates (indicating points in the video frames) to image-centered coordinates (referring to the stimulus images). 

After truncating the data to the relevant time segments of the stimulus presentation, we used a velocity-based saccade detection algorithm \citep{Engbert2003,Engbert2006}. For more detailed information on how to fit the parameters to our measurement device, please see \citet{Backhaus2019}, where a number of filter criteria are described in detail. These criteria  produce reliable data points, when working with the SMI Eye Tracking Glasses (SMI-ETG 2W; SensoMotoric Instruments, Teltow, Germany). After preprocessing, a total of 40,182 fixations and 47,425 saccades were retained for further analyses and modeling.

\subsection{Most important results}
The original experiment by \citet{Backhaus2019} reports statistics, from which we summarize the most relevant effects in the following. The authors looked at temporal and spatial eye movement parameters and compared the 4 different tasks using linear mixed models. The contrasts of the linear mixed models were chosen in such a way that the differences between the task groups (Guess conditions/free viewing vs.~Count conditions/search) as well as the differences between the two specific tasks within a type could be compared (Guess time vs.~Guess country; Count people vs.~Count animals).

The authors report variations in fixation durations induced by the experimental task manipulations. On average, fixation durations are shorter in Count tasks compared to Guess tasks. Particularly short fixation durations occur in difficult Count tasks; Couting animals involves more challenging search components than counting people. Results also showed differences in saccade amplitudes between task types: Count tasks lead to shorter saccade amplitudes than Guess tasks.  For saccade amplitudes, unlike fixation durations, no differences were found within task types. \citet{Backhaus2019} report that the tasks produced differences in gaze behavior on other spatial parameters. In Count tasks, participants disengaged faster and further from the image center (after generating the initial tendency to fixate the image center) compared to Guess conditions \citep{Rothkegel2017, Tatler2007}. 

With respect to the image-dependent 2D density of fixations, gaze in the Count people condition focused on comparatively fewer salient locations while fixation locations in the Count animals condition were most distributed across the image. The Guess tasks induced distributions between these two extremes. Thus, there was a strong influence of the task on image-dependent saliency.

\newpage
\addcontentsline{toc}{section}{Appendix B: Model Equations}
\setcounter{table}{0}
\renewcommand{\thetable}{B\arabic{table}} 
\setcounter{figure}{0}
\renewcommand{\thefigure}{B\arabic{figure}} 
\setcounter{equation}{0}
\renewcommand{\theequation}{B\arabic{equation}} 

\section{SceneWalk Model Specification}
\label{App:Model}
In the main text, we introduced the basic components of the SceneWalk model in its most recent version \citep{Schwetlick2020}. We provide additional mathematical details in this appendix. As explained in the main text, the SceneWalk model comprises two largely independent processing streams, activation and inhibition, which when combined are interpreted as the fixation probability $\pi$ at each grid point $i, j$ at time $t$. In the original formulation of the model \citep{Engbert2015}, the center of both the activation and the inhibition stream align with the current fixation position ($f_x, f_y$). The differential equations that define the temporal evolution of the activations of the two streams are given in Eq.~(\ref{eq:diffA}) for the activation stream and in Eq.~(\ref{eq:diffF}) for the inhibitory stream in the main text.

Over time intervals with constant input (i.e., during fixation, a closed-form solution can be found by integrating analytically, i.e., for the activation 
\begin{equation}
    A(t) = \frac{G_AS}{\sum{G_AS}} + e^{-\omega_A(t-t_0)}\left(A_0 -\frac{G_AS}{\sum{G_AS}}\right),
    \label{eq:att_eq}
\end{equation}
and
\begin{equation}
    F(t) = \frac{G_F}{\sum{G_F}} + e^{-\omega_F(t-t_0)}\left(F_0 -\frac{G_F}{\sum{G_F}}\right),
    \label{eq:inhib_eq}
\end{equation}
for the inhibition, where we dropped the indices $i, j$ for simplification of the notation. It is important to note that the assumption of constant input is an approximation because of the presence of miniature eye movement produced involuntarily during fixation \citep[e.g.,][]{Engbert2006}.

The weighted difference of the activations in the two streams represents the priority map for target selection (Eq.~(\ref{eq:subtr}). Since the difference will lead negative activations at locations, we take the part of the map, i.e., 
\begin{equation}
\label{eq:norm}
    u^*_{ij}(u_{ij})= \left\{
    \begin{array}{cl}
      u_{ij}, & \mbox{if}\ u_{ij}>0 \\
      0, & \mbox{otherwise} .
    \end{array}\right. \;.
\end{equation}
The most recent version of the SceneWalk model \citep{Schwetlick2020} introduced different phases of perisaccadic influences during each fixation. Specifically, before and after a saccade, the center of the activation stream shifts. A pre-saccadic shift to the upcoming target occurs before saccade onset and post-saccadic shift in the direction of the saccade vector occurs after the saccade (Fig.~\ref{fig:phases}).  Thus, for a time $\tau_{pre}$ before each saccade, once the next location has been selected from the priority map with probability $\pi(i,j)$, the center of the Gaussian input shifts to the location of the upcoming fixation, i.e., 
\begin{equation}
\label{eq:preshift}
    G_{A}^{\rm pre}(x,y) = \frac{1}{2\pi \sigma_{A}^2} \exp\left(-\frac{(x-x_{f+1})^2+(y-y_{f+1})^2}{2\sigma_{A}^2}\right)  \;,
\end{equation}
When the pre-saccadic phase terminates, the saccade is executed. For the purposes of this work, we neglect saccade durations, as most information is acquired during fixations. Now, the post-saccadic shift phase begins, during which the center of the activation Gaussian is determined by Eq.~(\ref{eq:shiftloc2}). The evolution equation is then given by
\begin{equation}
    G_A^{post}(x,y) = \frac{1}{2\pi \sigma_{post}^2} \exp\left(-\frac{(x-x_{\rm s})^2+(y-y_{\rm s})^2}{2\sigma_{post}^2}\right) \;.
    \label{eq:post_gauss}
\end{equation}

As the inhibition stream always aligns with the fixation location, it can still be calculated for the entire fixation duration via Eq.~(\ref{eq:inhib_eq}). The result of the phase-specific activation and inhibition can be combined at any point in time to yield the fixation selection probability at that time.

Facilitation of return is implemented in the model as a selectively slower decay of attention $\omega_A$ at the one back location. It thus occurs more briefly and at a different time scale than the inhibition of return implemented in the inhibition stream. The reduced decay rate $\omega_{\rm FoR}$ occurs in a spatial window  $x-\nu<x_{f-1}<x+\nu$ and $y-\nu<y_{f-1}<y+\nu$ around the previous fixation location $(x_{f-1},y_{f-1})$, where $\nu$ is the size of the window. We then replace $\omega_A$ in the evolution equation with a matrix that contains the value of $\omega_A$ everywhere except in the specified window, where it contains $\omega_{\rm FoR}$ 
\begin{equation}
    A(t) = \frac{G_AS}{\sum{G_AS}} + e^{-\omega_{FoR}(t-t_0)}\left(A_0 -\frac{G_AS}{\sum{G_AS}}\right).
    \label{eq:att_om_for}
\end{equation}

As suggested by \citet{Rothkegel2017}, starting the model with a central activation improves the predictions of the model. 
Initially we instantiated the model with uniform distributions. 
The implementation of a transient central fixation bias changes the evolution equation for the first fixation so that 
\begin{equation}
    A(t) = \frac{G_{fix}S}{\sum{G_{fix}S}} + e^{-\omega_{cb}(t-t_0)}\left(A_{0_{CB}} -\frac{G_{fix}S}{\sum{G_{fix}S}}\right).
    \label{eq:cb_evo}
\end{equation}

Finally, we implemented an additional bias towards horizontal and verical saccade directions \citep{Engbert2011}. The oculomotor map is centered at the current fixation location, i.e., 
\begin{equation}
\label{eq:omp1}
    P_{OM} = \left((x-x_f)^2 \cdot (y-y_f)^2\right)^\chi \;,
\end{equation}
where the factor $\chi$ determines the steepness of the oculomotor potential. In this variation, before the normalization and the addition of noise, Eq.~(\ref{eq:norm}, \ref{eq:noise}), the oculomotor map is added as
\begin{equation}
\label{eq:omp2}
    u_{\rm OM} = u + \left(\psi \cdot \left|\frac{P_{OM}}{\max(P_{OM})}-1\right|\right) \;,
\end{equation}
where $\psi=10^{-0.6}$ is a constant parameter. 

\newpage

\addcontentsline{toc}{section}{Appendix C: Workflow on Bayesian inference}
\setcounter{table}{0}
\renewcommand{\thetable}{C\arabic{table}} 
\setcounter{figure}{0}
\renewcommand{\thefigure}{C\arabic{figure}} 
\setcounter{equation}{0}
\renewcommand{\theequation}{C\arabic{equation}} 
\section{Bayesian Inference Workflow}
\label{App:Info}
\begin{figure}[p]
\caption{\label{fig:infographic}
Workflow for likelihood-based Bayesian inference.}
\includegraphics[angle=90,width=120mm]{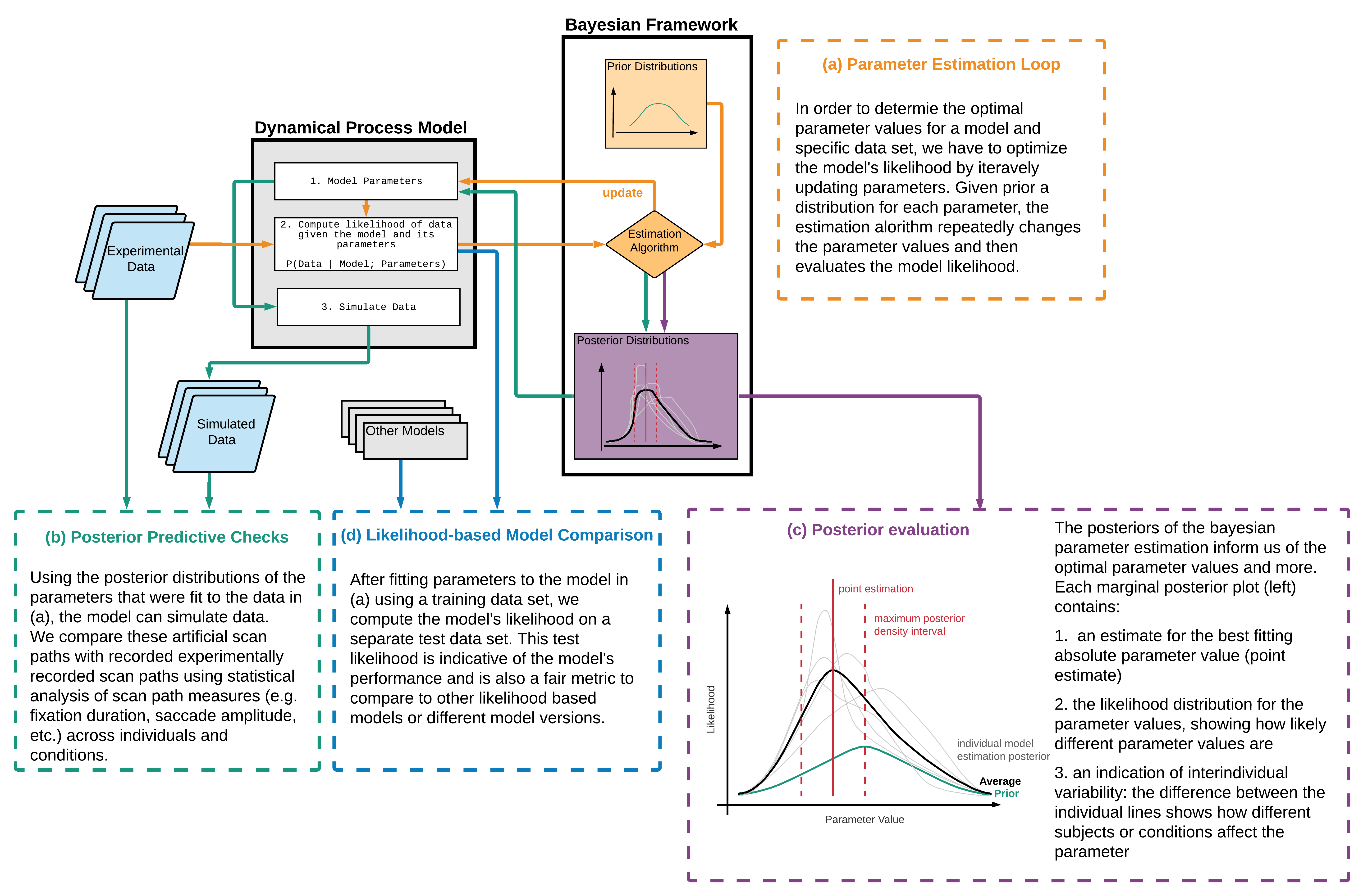}
\figurenote{The workflow summarizes all steps of Bayesian inference and highlights four core advantages of the proposed workflow. Orange arrows and lines (a) refer to the statistically rigorous estimation of parameters using the model's likelihood function and empirical data. Green arrows and lines (b) show the process of conducting posterior predictive checks, where the resultant models' predictions are evaluated against real world data. Purple arrows and lines (c) explain how the specific parameter posteriors can be interpreted in a biologically-founded model. Lastly, blue arrows and lines (d) explain how the method is useful to establish comparability between competing models.}
\end{figure}

In this paper we applied a Bayesian inference workflow to a biologically plausible generative model. This approach is extremely promising for cognitive modeling for four reasons illustrated in the infographic in Fig. \ref{fig:infographic}. 

In this framework a model is defined by its likelihood function and parameters. It can be used to calculate the probability of a given data point. Given a starting point it can also be used generatively to simulate data. Both the predictive and the generative parts of the model are necessary components of the proposed workflow and provide valuable insight into the model's characteristics.

First, we use the model likelihood to estimate the best values for the model parameters using Bayesian inference. The Bayesian parameter estimation algorithm repeatedly computes the model likelihood given the data, while systematically varying the parameter values. Thus, it tries to maximize the performance of the model using the likelihood given the data. This process yields marginal posterior distributions for each parameter. These marginal posteriors can be interpreted as a rich source of information about the parameter as shown in box (c). 

Second, we parametrize the model with the values obtained from the estimation and use it to simulate data. When fitting a model using an ad-hoc loss function, the model is trained specifically to reproduce whatever the chosen metrics may be. By contrast, using the likelihood allows for greater generalizability as well as avoiding overfitting. Simulated data can be compared to experimental data in order to assess how well the model reproduced trends that it was not directly informed of. To this end we perform a series of posterior predictive checks, which ascertain whether the model can actually capture the relevant features found in the data. Thus, they reveal strengths and weaknesses of the model regarding its plausibility.

Lastly, the model likelihood is relevant also for inter-model comparisons. It is a fair basis for comparison, in the sense that it provides the same information to each model with the experimental data. Each model can be fitted and compared in the same way: estimation algorithms determine the parameters using a training set of experimental data. Then, using a test set of experimental data, we can calculate and compare their performance.

\newpage
\addcontentsline{toc}{section}{Appendix D: Convergence of Parameter Estimation}
\setcounter{table}{0}
\renewcommand{\thetable}{D\arabic{table}} 
\setcounter{figure}{0}
\renewcommand{\thefigure}{D\arabic{figure}} 
\setcounter{equation}{0}
\renewcommand{\theequation}{D\arabic{equation}} 

\section{Convergence of Parameter Estimation}
\label{sec:appx_conv}
\begin{figure}[p]
\caption{\label{fig:conv}
Rhat Convergence}
\centering
\includegraphics[angle=0,width=120mm]{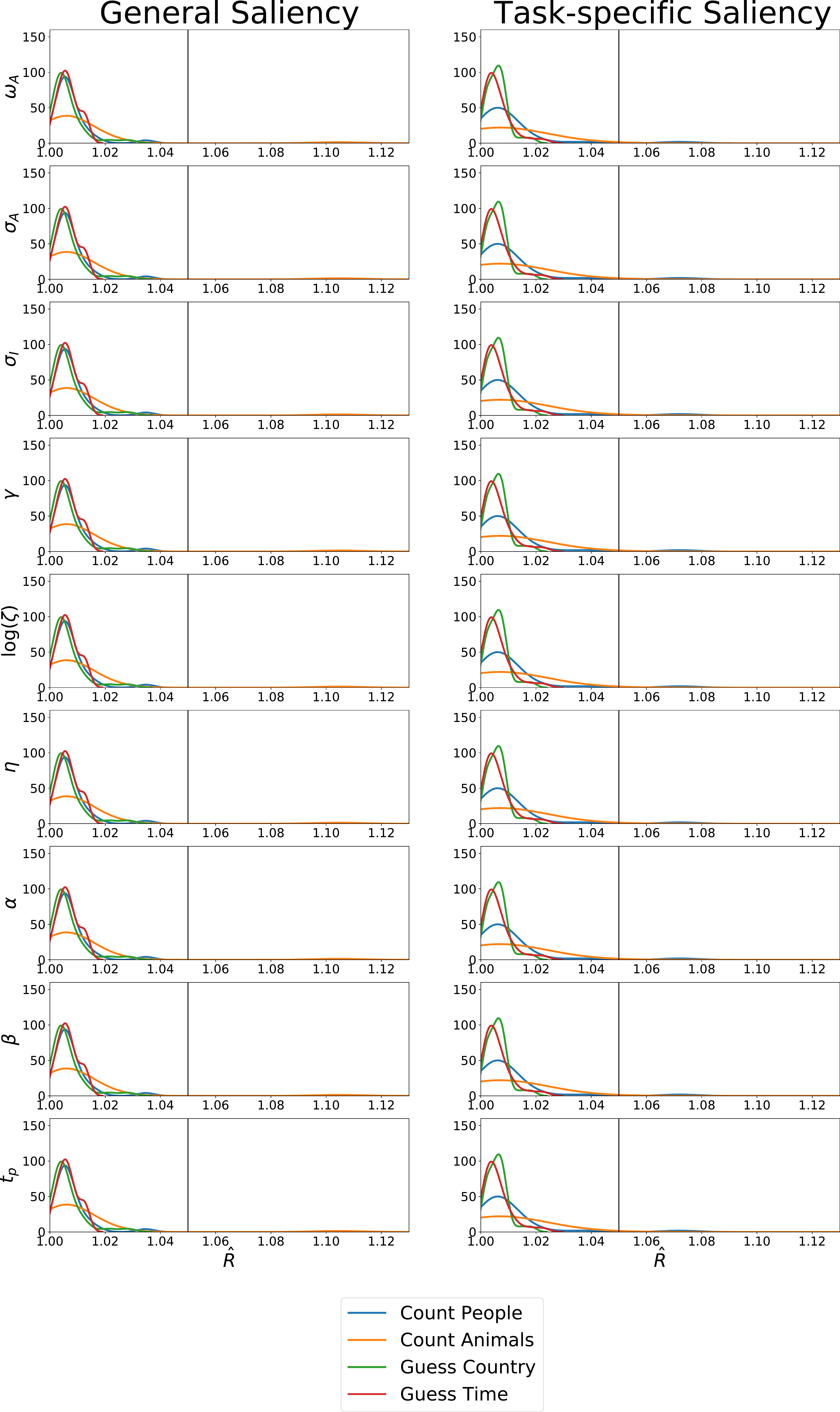}
\figurenote{The plot shows the distribution of $\hat{R}$ values over our models. Each panel row represent a parameter. The horizontal line indicates the value of 1.05, a common threshold for determining convergence.
}
\end{figure}

As suggested by the authors of the DREAM algorithm \citep{Vrugt2011}, we used the Gelman-Rubin convergence diagnostic $\hat{R}$ to determine adequate quiality of the parameter estimation. The results are illustrated below for for all 256 models (Figure \ref{fig:conv}). We used the value of 1.05 as a threshold to indicate convergence. In total of the 2304 fitted parameters, 2288 converged and 16 did not converge. At the level of models, of the 256 fitted models there were only 3 where the posterior did not converge for one or more parameters.

\newpage
\section{Preregistration}
This work was preregistered at the Open Science Framework (OSF)\footnote{See \url{https://osf.io/79qy8}} \citep{Schwetlick_prereg} using the \enquote{Preregistration Template for the Application of Cognitive Models} \citep{Cruewell2019}. Please refer to the OSF repository for full information on the preregistration. Here we would like to follow up on some aspects of the preregistration and explain where and why we deviated from the preregistered research plan.

The hypotheses we stated in the preregistration concerned (a) differences in model parameters relating to the attention span for the different tasks and (b) the importance of inhibition for different tasks. For the former hypothesis our findings agreed that the attentional span is greater in Guess task conditions than in Count tasks. As predicted, the activation Gaussian $\sigma_A$ is greater in free viewing-like tasks.

We also find support for the latter hypothesis: In the Count task conditions, the span for the fixation map $\sigma_F$ is smaller than in the Guess tasks, showing at least a more focused, localized inhibition component. The parameter $C_F$, which is mentioned as an exploratory analysis target in the preregistration was not included as a free parameter in the final estimation, as it turned out to be more difficult to identify given the relatively small amount of data available for each model fit. 

The third hypothesis in the preregistration concerns parameter $\omega_A$, which controls the speed of decay. We predicted a smaller value of $\omega_A$ for Count tasks, as we thought that keeping track of past fixations would be of greater use. We found this to be true, but only for models fitted using the general saliency map, not for models using task-specific saliency. We propose potential reasons for this finding in the discussion. 

We also proposed a Markov-order analysis of the model to determine the influence of past states on the current. This analysis is not included in the current manuscript, since pilot simulations indicated that the analysis required larger amounts of data per fit than available from the current study. However, we consider the mathematical concept promising and aim to include a corresponding analysis in future work. The same is true for the mean-lag distance analysis proposed in the preregistration.

An important point in the preregistration states was the possibility of running model fits based on individual data sets per task. As this was successful despite the limited data, the results of the current work are exclusively based on this strategy of fitting data for individuals and tasks independently. The alternative proposal of fitting models for each task by pooling across participants was no longer necessary. Additionally, instead of the proposed 5 free parameters for model fitting, we now successfully estimated 9 free parameters per data set, with 3 free parameters added the model to include the new temporal control of fixation durations.

\newpage
\addcontentsline{toc}{section}{Appendix E: Additional Results}
\setcounter{table}{0}
\renewcommand{\thetable}{F\arabic{table}} 
\setcounter{figure}{0}
\renewcommand{\thefigure}{F\arabic{figure}} 
\setcounter{equation}{0}
\renewcommand{\theequation}{F\arabic{equation}} 
\section{Additional Results}
The following section provides additional details concerning the statistical analyses presented in this paper. Specifically, we provide the detailed results of our LMM analyses. Table \ref{ex:lmm_structure} summarizes all of the applied LMM model structures.

The first LMM analysis compares the likelihoods of different versions of the SceneWalk model and Density Sampling models. The results can be found in Table \ref{ex:LMM_loglik}. The second analysis comprises an LMM of the posterior of each of the 9 estimated parameters for both general and task-specific model variants, i.e., 18 separate models. Fixed and random effects for each LMM are reported in Table \ref{ex:LMM_modelparam}. Finally, Tables \ref{ex:LMMSacAmpl} and \ref{ex:LMMFixDur} contain the results of the LMMs pertaining to the comparison of simulated and experimental saccade amplitudes and fixation durations, resp.

\begin{table}[h]
\begin{center}
\color{black}
\caption{LMM fit by maximum likelihood - Comparison of the model likelihood gain.}
\input{TabF1.txt}

\figurenote{$|t| > 2$ are interpreted as significant effects, FModel: factor model 'Density Sampling' vs. 'SceneWalk', FSal: factor saliency 'General Saliency' vs. 'Task-specific Saliency', FInter: Factor for the interaction of FModel and FSal, 
T1: 'Task-specific Saliency - Density Sampling', 
T2: 'Task-specific Saliency - SceneWalk',
T3: 'General Saliency - Density Sampling',
T4: 'General Saliency - Scene Walk', To avoid zeros in the model likelihood gain, data were linearly transformed by adding 14.}
\label{ex:LMM_loglik}
\end{center}
\end{table}

\begin{table}[p]
\begin{center}
\color{black}
\caption{LMM fit by maximum likelihood - Model parameter with our custom contrasts.}

\input{TabF2_1.txt}
\figurenote{FGC: first contrast 'Count' vs. 'Guess', FC: second contrast 'Count People' vs. 'Count Animals', FG: third contrast 'Guess Country' vs. 'Guess Time', $|t| > 2$ are interpreted as significant effects.}
\label{ex:LMM_modelparam}
\end{center}
\end{table}

\renewcommand{\thetable}{F2 (cont'd)}
\begin{table}[p]
\begin{center}
\color{black}
\caption{LMM fit by maximum likelihood - Model parameter with our custom contrasts.}

\input{TabF2_2.txt}
\figurenote{FGC: first contrast 'Count' vs. 'Guess', FC: second contrast 'Count People' vs. 'Count Animals', FG: third contrast 'Guess Country' vs. 'Guess Time', $|t| > 2$ are interpreted as significant effects.}\end{center}
\end{table}

\renewcommand{\thetable}{F2 (cont'd)}
\begin{table}[p]
\begin{center}
\color{black}
\caption{LMM fit by maximum likelihood - Model parameter with our custom contrasts.}

\input{TabF2_3.txt}
\figurenote{FGC: first contrast 'Count' vs. 'Guess', FC: second contrast 'Count People' vs. 'Count Animals', FG: third contrast 'Guess Country' vs. 'Guess Time', $|t| > 2$ are interpreted as significant effects.}\end{center}
\end{table}

\setcounter{table}{2}
\renewcommand{\thetable}{F\arabic{table}}
\begin{table}[p]
\begin{center}
\color{black}
\caption{LMM fit by maximum likelihood -- Saccade amplitudes (log-transformed) for our contrasts.}
\rotatebox{90}{
\input{TabF3.txt}
}
\figurenote{$|t| > 2$ are interpreted as significant effects.}
\label{ex:LMMSacAmpl}
\end{center}
\end{table}

\begin{table}[p]
\begin{center}
\color{black}
\caption{LMM fit by maximum likelihood -- Fixation durations (log-transformed) for our contrasts.}
\rotatebox{90}{
\input{TabF4.txt}
}
\figurenote{$|t| > 2$ are interpreted as significant effects.}
\label{ex:LMMFixDur}
\end{center}
\end{table}

\begin{table}[p]
\begin{center}
\color{black}

\caption{LMM model structure.}
\input{TabF5.txt}

\figurenote{1: Intercept, FModel: factor model 'Density Sampling' vs. 'SceneWalk', FSal: factor saliency 'General Saliency' vs. 'Task-specific Saliency', FInter: Factor for the interaction of FModel and FSal, 
T1: 'Task-specific Saliency - Density Sampling', 
T2: T1 vs. 'Task-specific Saliency - SceneWalk',
T3: T1 vs. 'General Saliency - Density Sampling',
T4: T1 vs. 'General Saliency - Scene Walk',
FGC: first contrast 'Count' vs. 'Guess', FC: second contrast 'Count People' vs. 'Count Animals', FG: third contrast 'Guess Country' vs. 'Guess Time', ||: double bar sign represents that the correlations of random effects are not included in the model, *we choose the minimal model with only random intercepts for subjects and images to have comparable models between all subsets of this analysis.} 
\label{ex:lmm_structure}

\end{center}
\end{table}

\end{appendix}
\end{document}

%% file: Tab1.txt
\begin{tabular}{ c l c r@{ ... }l c c } 
\toprule
Parameter        & Description                                                & Eq.                  & \multicolumn{2}{c}{Range}          & Mean               & SD \\ 
\midrule                                                                                             
$\omega_A$       & Speed of decay of the activation stream                    & (\ref{eq:diffA})     & 0                &       100   & 10                 &  12  \\ 
$\sigma_A$       & Standard deviation of the Gaussian activation (°)          & (\ref{eq:diffA})     & 0                &      30     & 7                  &  5    \\ 
$\sigma_F$       & Standard deviation of the Gaussian inhibition (°)          & (\ref{eq:diffF})     & 0                &      30     & 4                  &  4    \\ 
$\gamma_i$       & Exponent regulating determinism in target selection        & (\ref{eq:subtr})     & 0                &      5      & 1                  &  3   \\ 
$\log_{10}\zeta$ &  Noise parameter for target selection                      & (\ref{eq:noise})     & $-10$            &      0      & $-2$               &  2   \\ 
$\eta$           & Size of the post-saccadic shift relative to saccade length & (\ref{eq:shiftloc2}) & 0                &      4      & 0.5                &  2    \\ 
$t_\alpha$       & Timing intercept                                           & (\ref{eq:parb})      & 0                &      5      & 3                  &  5    \\ 
$t_\beta$        & Factor for the coupling of saliency and timer              & (\ref{eq:parb})      & $-4$             &      0      & $-0.4$             &  3    \\ 
$q$              & Shape parameter for the timing distribution                & (\ref{eq:gamma})     & 0                &      15     & 3                  &  3     \\
\bottomrule
\end{tabular}

%% file: Tab2.txt
\begin{tabular}{llcccccccc}
\toprule
                  Model &     Parameter & \multicolumn{2}{c}{Count Animals} & \multicolumn{2}{c}{Count People} & \multicolumn{2}{c}{Guess Country} & \multicolumn{2}{c}{Guess Time} \\
                        &               & +/- & Point Estimate &          +/- & Point Estimate &           +/- & Point Estimate &        +/- & Point Estimate \\
\midrule
       General Saliency &      $\gamma$ &         0.099 &          0.913 &        0.081 &          1.053 &         0.143 &          0.932 &      0.091 &          0.938 \\
       General Saliency &    $\omega_A$ &         2.457 &          8.378 &        2.863 &          8.918 &         4.774 &         12.361 &      4.280 &         10.927 \\
       General Saliency &        $\eta$ &         0.083 &          0.253 &        0.060 &          0.179 &         0.089 &          0.170 &      0.084 &          0.182 \\
       General Saliency &    $\sigma_A$ &         1.710 &          6.913 &        1.744 &          9.005 &         1.936 &         10.506 &      1.632 &         10.224 \\
       General Saliency &    $\sigma_F$ &         1.292 &          5.049 &        1.790 &          8.146 &         2.523 &          8.437 &      1.911 &          9.116 \\
       General Saliency &  $t_{\alpha}$ &         1.036 &          3.964 &        1.185 &          3.815 &         1.171 &          3.829 &      1.019 &          3.981 \\
       General Saliency &   $t_{\beta}$ &         0.220 &         -1.350 &        0.174 &         -1.109 &         0.240 &         -1.352 &      0.259 &         -1.307 \\
       General Saliency &         $q$ &         0.498 &          5.053 &        0.477 &          4.475 &         0.618 &          4.757 &      0.664 &          5.010 \\
       General Saliency &  $log(\zeta)$ &         0.415 &         -1.122 &        0.492 &         -1.208 &         0.842 &         -1.861 &      0.738 &         -1.998 \\
 Task-specific Saliency &      $\gamma$ &         0.088 &          0.961 &        0.100 &          1.027 &         0.109 &          0.943 &      0.087 &          0.955 \\
 Task-specific Saliency &    $\omega_A$ &         3.080 &          9.680 &        4.187 &         11.815 &         4.642 &         13.129 &      3.779 &         11.184 \\
 Task-specific Saliency &        $\eta$ &         0.077 &          0.259 &        0.067 &          0.177 &         0.089 &          0.174 &      0.080 &          0.181 \\
 Task-specific Saliency &    $\sigma_A$ &         1.674 &          7.832 &        1.814 &          9.573 &         1.903 &         11.060 &      1.733 &         10.608 \\
 Task-specific Saliency &    $\sigma_F$ &         1.573 &          5.387 &        1.970 &          7.910 &         2.058 &          8.287 &      1.916 &          9.283 \\
 Task-specific Saliency &  $t_{\alpha}$ &         1.101 &          1.101 &        1.015 &          1.015 &         1.074 &          3.926 &      1.037 &          3.963 \\
 Task-specific Saliency &   $t_{\beta}$ &         0.226 &         -1.517 &        0.185 &         -1.280 &         0.248 &         -1.333 &      0.261 &         -1.307 \\
 Task-specific Saliency &         $q$ &         0.511 &          5.222 &        0.501 &          4.701 &         0.621 &          4.786 &      0.665 &          5.047 \\
 Task-specific Saliency &  $log(\zeta)$ &         0.613 &         -1.273 &        0.759 &         -1.585 &         0.826 &         -2.637 &      0.767 &         -2.622 \\
\bottomrule
\end{tabular}

%% file: TabF1.txt

\begin{tabular}{l l r@{.}l r@{.}l r@{.}l c  c  l l  r@{.}l r@{.}l r@{.}l }
\noalign{\vskip0.3cm}
\toprule
\multicolumn{2}{l}{}                       & \multicolumn{6}{l}{Custom contrast}                                               & \multicolumn{2}{l}{}      & \multicolumn{2}{l}{}          & \multicolumn{6}{l}{Treatment contrast}                                                 \tabularnewline
\hline                                                                                                                                                                                   
\multicolumn{18}{l}{}                                                                                                                                                                                           \tabularnewline
\multicolumn{2}{l}{}                       & \multicolumn{2}{c}{$\beta$} & \multicolumn{2}{c}{$SE$} & \multicolumn{2}{c}{$t$}  & \multicolumn{2}{l}{}      & \multicolumn{2}{l}{}          & \multicolumn{2}{c}{$\beta$}  & \multicolumn{2}{c}{$SE$}  & \multicolumn{2}{c}{$t$}     \tabularnewline
                                             \cline{3-8}                                                                                                                                     \cline{13-18}
\multicolumn{2}{l}{\textit{Fixed Effects}} & \multicolumn{16}{c}{}\tabularnewline

\multicolumn{2}{l}{Intercept}              & 15&10                       & 0&066                    &   228&74                 & \multicolumn{2}{l}{} &     \multicolumn{2}{r}{T1}        &  15&07                        & 0&069                     & 218&72                       \tabularnewline
\multicolumn{2}{l}{FModel}                 &  0&34                       & 0&023                    &    15&01                 & \multicolumn{2}{l}{} &     \multicolumn{2}{r}{T2 - T1}   &   0&31                        & 0&032                     &   9&72                       \tabularnewline
\multicolumn{2}{l}{FSal}                   &  0&25                       & 0&023                    &    11&04                 & \multicolumn{2}{l}{} &     \multicolumn{2}{r}{T3 - T1}   &  -0&28                        & 0&032                     &  -8&70                       \tabularnewline
\multicolumn{2}{l}{FInter}                 & -0&06                       & 0&046                    &    -1&26                 & \multicolumn{2}{l}{} &     \multicolumn{2}{r}{T4 - T1}   &   0&09                        & 0&032                     &   2&80                       \tabularnewline

\multicolumn{18}{l}{} \tabularnewline
\multicolumn{2}{l}{}                      & \multicolumn{2}{c}{$Var$}  & \multicolumn{2}{c}{$SD$} & \multicolumn{2}{c}{}       & \multicolumn{2}{l}{}      & \multicolumn{2}{l}{}          & \multicolumn{2}{c}{$Var$}  & \multicolumn{2}{c}{$SD$}  & \multicolumn{2}{c}{}     \tabularnewline
                                          \cline{3-8}                                                                                                                                     \cline{13-18}
\multicolumn{2}{l}{\textit{Random Effects}} & \multicolumn{16}{c}{}\tabularnewline
\multicolumn{2}{l}{Subject: Intercept}      & 0&0327                     & 0&181                  &   \multicolumn{2}{l}{}   & \multicolumn{2}{l}{}   & \multicolumn{2}{r}{}            &  0&0327                        & 0&181                    & \multicolumn{2}{l}{}         \tabularnewline
\multicolumn{2}{l}{Image: Intercept}        & 0&0960                     & 0&310                  &   \multicolumn{2}{l}{}   & \multicolumn{2}{l}{}   & \multicolumn{2}{r}{}            &  0&0960                        & 0&310                    & \multicolumn{2}{l}{}         \tabularnewline
\multicolumn{2}{l}{Residual}                & 0&2205                     & 0&470                  &   \multicolumn{2}{l}{}   & \multicolumn{2}{l}{}   & \multicolumn{2}{r}{}            &  0&2205                        & 0&470                    & \multicolumn{2}{l}{}         \tabularnewline
                                            \cline{3-8}                                                                                                                                 \cline{13-18}
\multicolumn{18}{l}{} \tabularnewline

\multicolumn{2}{l}{Number of obs}          & \multicolumn{2}{l}{1700}     & \multicolumn{2}{l}{}    &   \multicolumn{2}{l}{}   & \multicolumn{2}{l}{}      & \multicolumn{2}{l}{}         &  \multicolumn{2}{l}{1700}     & \multicolumn{2}{l}{}      & \multicolumn{2}{l}{}         \tabularnewline
\multicolumn{2}{l}{Number of groups}       & \multicolumn{2}{l}{subject}  & \multicolumn{2}{l}{32}  &   \multicolumn{2}{l}{}   & \multicolumn{2}{l}{}      & \multicolumn{2}{l}{}         &  \multicolumn{2}{l}{subject}  & \multicolumn{2}{l}{32}    & \multicolumn{2}{l}{}         \tabularnewline
\multicolumn{2}{l}{}                       & \multicolumn{2}{l}{image}    & \multicolumn{2}{l}{30}  &   \multicolumn{2}{l}{}   & \multicolumn{2}{l}{}      & \multicolumn{2}{l}{}         &  \multicolumn{2}{l}{image}    & \multicolumn{2}{l}{30}    & \multicolumn{2}{l}{}         \tabularnewline
\multicolumn{18}{l}{}                                                                                                                                                                                           \tabularnewline
\bottomrule

\end{tabular}\\

%% file: TabF2_1.txt
\begin{tabular}{lccclllccc}
\toprule

                   & \multicolumn{3}{l}{$\omega_A$ -- Task-specific Saliency}         &  &  &  & \multicolumn{3}{l}{$\omega_A$ -- General Saliency}          \\
\textit{Fixed Effects}      & $\beta$           & $SE$                     & $t$      &  &  &  & $\beta$           & $SE$               & $t$      \\
\cline{2-4} \cline{8-10}
Intercept          &  0.760         & 0.004                  & 204.85  &  &  &  &  1.053         & 0.001            & 1102.15 \\
FGC                & -0.001         & 0.006                  &  -0.22  &  &  &  &  0.006         & 0.002            &    2.60 \\
FC                 & -0.014         & 0.011                  &  -1.20  &  &  &  &  0.000         & 0.002            &   -0.07 \\
FG                 &  0.013         & 0.007                  &   2.03  &  &  &  & -0.006         & 0.002            &   -2.59 \\
\textit{Random Effects}    & $Var$            & $SD$                   &         &  &  &  & $Var$            & $SD$               &         \\
\cline{2-4} \cline{8-10}
Subject: Intercept & 0.0004         & 0.021                  &         &  &  &  & 0.0000         & 0.005            &         \\
Subject: FGC       & 0.0011         & 0.033                  &         &  &  &  & 0.0001         & 0.012            &         \\
Subject: FC        & 0.0041         & 0.064                  &         &  &  &  & 0.0002         & 0.014            &         \\
Subject: FG        & 0.0014         & 0.037                  &         &  &  &  & 0.0002         & 0.014            &         \\
Residual           & 0.0005         & 0.022                  &         &  &  &  & 0.0000         & 0.006            &         \\
                   &                &                        &         &  &  &  &                &                  &         \\
                   & \multicolumn{3}{l}{$\sigma_A$ -- Task-specific Saliency}         &  &  &  & \multicolumn{3}{l}{$\sigma_A$ -- General Saliency}        \\
\textit{Fixed Effects}      & $\beta$           & $SE$                     & $t$      &  &  &  & $\beta$           & $SE$               & $t$      \\
\cline{2-4} \cline{8-10}
Intercept          & 2.923          & 0.037                  & 79.17   &  &  &  & 5.381          & 0.119            & 45.11   \\
FGC                & 0.233          & 0.062                  &  3.79   &  &  &  & 0.760          & 0.209            &  3.63   \\
FC                 & 0.139          & 0.076                  &  1.82   &  &  &  & 0.682          & 0.226            &  3.01   \\
FG                 & 0.022          & 0.058                  &  0.38   &  &  &  & 0.154          & 0.234            &  0.66   \\
\textit{Random Effects}    & $Var$            & $SD$                   &         &  &  &  & $Var$            & $SD$               &         \\
\cline{2-4} \cline{8-10}
Subject: Intercept & 0.0436         & 0.209                  &         &  &  &  & 0.4551         & 0.675            &         \\
Subject: FGC       & 0.1214         & 0.348                  &         &  &  &  & 1.4028         & 1.184            &         \\
Subject: FC        & 0.1852         & 0.430                  &         &  &  &  & 1.6362         & 1.279            &         \\
Subject: FG        & 0.1085         & 0.329                  &         &  &  &  & 1.7575         & 1.326            &         \\
Residual           & 0.0352         & 0.188                  &         &  &  &  & 0.3657         & 0.605            &         \\
                   &                &                        &         &  &  &  &                &                  &         \\
                   & \multicolumn{3}{l}{$\sigma_F$     -- Task-specific Saliency}     &  &  &  & \multicolumn{3}{l}{$\sigma_F$     -- General Saliency}        \\
\textit{Fixed Effects}      & $\beta$           & $SE$                     & $t$      &  &  &  & $\beta$           & $SE$               & $t$      \\
\cline{2-4} \cline{8-10}
Intercept          & 4.035          & 0.069                  & 58.40   &  &  &  & 4.710          & 0.103            & 45.65   \\
FGC                & 0.626          & 0.132                  &  4.75   &  &  &  & 0.778          & 0.164            &  4.73   \\
FC                 & 0.449          & 0.169                  &  2.65   &  &  &  & 0.836          & 0.211            &  3.97   \\
FG                 & 0.217          & 0.181                  &  1.20   &  &  &  & 0.448          & 0.245            &  1.83   \\
\textit{Random Effects}    & $Var$            & $SD$                   &         &  &  &  & $Var$            & $SD$               &         \\
\cline{2-4} \cline{8-10}
Subject: Intercept & 0.1526         & 0.391                  &         &  &  &  & 0.3401         & 0.583            &         \\
Subject: FGC       & 0.5550         & 0.745                  &         &  &  &  & 0.8629         & 0.929            &         \\
Subject: FC        & 0.9145         & 0.956                  &         &  &  &  & 1.4193         & 1.191            &         \\
Subject: FG        & 1.0450         & 1.022                  &         &  &  &  & 1.9103         & 1.382            &         \\
Residual           & 0.3575         & 0.598                  &         &  &  &  & 0.6995         & 0.836            &         \\
                   &                &                        &         &  &  &  &                &                  &         \\
 
\bottomrule
\end{tabular}

%% file: TabF2_2.txt
\begin{tabular}{lccclllccc}
\toprule

                   & \multicolumn{3}{l}{$\gamma_i$          -- Task-specific Saliency} &  &  &  & \multicolumn{3}{l}{$\gamma_i$          -- General Saliency}        \\
\textit{Fixed Effects}      & $\beta$           & $SE$                     & $t$      &  &  &  & $\beta$           & $SE$               & $t$      \\
\cline{2-4} \cline{8-10}
Intercept          & 1.000          & 0.000                  & 3338.22 &  &  &  &  0.996         & 0.005            & 183.68  \\
FGC                & 0.001          & 0.001                  &    2.38 &  &  &  & -0.005         & 0.010            &  -0.49   \\
FC                 & 0.000          & 0.001                  &   -0.41 &  &  &  &  0.035         & 0.009            &   4.02    \\
FG                 & 0.000          & 0.001                  &   -0.57 &  &  &  & -0.003         & 0.016            &  -0.17   \\
\textit{Random Effects}    & $Var$            & $SD$                     &         &  &  &  & $Var$            & $SD$               &         \\
\cline{2-4} \cline{8-10}
Subject: Intercept & 0.0000         & 0.002                  &         &  &  &  & 0.0009         & 0.031            &         \\
Subject: FGC       & 0.0000         & 0.003                  &         &  &  &  & 0.0031         & 0.056            &         \\
Subject: FC        & 0.0000         & 0.005                  &         &  &  &  & 0.0024         & 0.049            &         \\
Subject: FG        & 0.0000         & 0.004                  &         &  &  &  & 0.0085         & 0.092            &         \\
Residual           & 0.0000         & 0.002                  &         &  &  &  & 0.0016         & 0.039            &         \\
                   &                &                        &         &  &  &  &                &                  &         \\
                   & \multicolumn{3}{l}{$\log_{10}\zeta$           -- Task-specific Saliency}  &  &  &  & \multicolumn{3}{l}{$\log_{10}\zeta$           -- General Saliency}      \\
\textit{\textit{Fixed Effects}}      & $\beta$           & $SE$                     & $t$      &  &  &  & $\beta$           & $SE$               & $t$      \\
\cline{2-4} \cline{8-10}
Intercept          & 31.770         & 0.455                  & 69.77   &  &  &  &  61.388        & 0.856            & 71.72   \\
FGC                & -4.502         & 0.980                  & -4.60   &  &  &  & -11.040        & 2.398            & -4.60   \\
FC                 & -2.098         & 1.527                  & -1.37   &  &  &  &  -3.719        & 3.014            & -1.23   \\
FG                 & -0.858         & 0.960                  & -0.89   &  &  &  &  -0.482        & 2.758            & -0.17   \\
\textit{Random Effects}    & $Var$            & $SD$                     &         &  &  &  & $Var$            & $SD$               &         \\
\cline{2-4} \cline{8-10}
Subject: Intercept &  6.6174        & 2.572                  &         &  &  &  &  23.3713       &  4.834           &         \\
Subject: FGC       & 30.6415        & 5.535                  &         &  &  &  & 183.7608       & 13.556           &         \\
Subject: FC        & 74.5076        & 8.632                  &         &  &  &  & 290.1551       & 17.034           &         \\
Subject: FG        & 29.3677        & 5.419                  &         &  &  &  & 242.8092       & 15.582           &         \\
Residual           & 32.3881        & 5.691                  &         &  &  &  & 125.8949       & 11.220           &         \\
                   &                &                        &         &  &  &  &                &                  &         \\
                   & \multicolumn{3}{l}{$\eta$    -- Task-specific Saliency}          &  &  &  & \multicolumn{3}{l}{$\eta$    -- General Saliency}         \\
\textit{Fixed Effects}      & $\beta$           & $SE$                     & $t$      &  &  &  & $\beta$           & $SE$               & $t$      \\
\cline{2-4} \cline{8-10}
Intercept          &  0.861         & 0.006                  & 155.14  &  &  &  &  0.808         & 0.007            & 112.35  \\
FGC                & -0.005         & 0.013                  &  -0.42  &  &  &  & -0.004         & 0.016            & -0.22   \\
FC                 & -0.017         & 0.013                  &  -1.29  &  &  &  & -0.032         & 0.017            & -1.92   \\
FG                 &  0.002         & 0.011                  &   0.19  &  &  &  & -0.001         & 0.016            & -0.04   \\
\textit{Random Effects}    & $Var$            & $SD$                   &         &  &  &  & $Var$            & $SD$               &         \\
\cline{2-4} \cline{8-10}
Subject: Intercept & 0.0010         & 0.031                  &         &  &  &  & 0.0017         & 0.041            &         \\
Subject: FGC       & 0.0052         & 0.072                  &         &  &  &  & 0.0085         & 0.092            &         \\
Subject: FC        & 0.0055         & 0.074                  &         &  &  &  & 0.0090         & 0.095            &         \\
Subject: FG        & 0.0038         & 0.062                  &         &  &  &  & 0.0078         & 0.088            &         \\
Residual           & 0.0013         & 0.035                  &         &  &  &  & 0.0018         & 0.042            &         \\
                   &                &                        &         &  &  &  &                &                  &         \\

\bottomrule
\end{tabular}

%% file: TabF2_3.txt
\begin{tabular}{lccclllccc}
\toprule
                   & \multicolumn{3}{l}{$t_\alpha$       -- Task-specific Saliency}   &  &  &  & \multicolumn{3}{l}{$t_\alpha$       -- General Saliency}         \\
\textit{Fixed Effects}      & $\beta$           & $SE$                     & $t$      &  &  &  & $\beta$           & $SE$               & $t$      \\
\cline{2-4} \cline{8-10}
Intercept          &  1.749         & 0.022                  & 80.28   &  &  &  &  2.108         & 0.026            & 81.29   \\
FGC                &  0.272         & 0.044                  &  6.18   &  &  &  &  0.024         & 0.060            &  0.39   \\
FC                 & -0.067         & 0.060                  & -1.12   &  &  &  & -0.113         & 0.080            & -1.41   \\
FG                 &  0.022         & 0.052                  &  0.42   &  &  &  &  0.127         & 0.069            &  1.85   \\
\textit{Random Effects}    & $Var$            & $SD$                   &         &  &  &  & $Var$            & $SD$               &         \\
\cline{2-4} \cline{8-10}
Subject: Intercept & 0.0149         & 0.122                  &         &  &  &  & 0.0210         & 0.145            &         \\
Subject: FGC       & 0.0606         & 0.246                  &         &  &  &  & 0.1121         & 0.335            &         \\
Subject: FC        & 0.1132         & 0.336                  &         &  &  &  & 0.2013         & 0.449            &         \\
Subject: FG        & 0.0836         & 0.289                  &         &  &  &  & 0.1473         & 0.384            &         \\
Residual           & 0.5813         & 0.762                  &         &  &  &  & 0.9381         & 0.969            &         \\
                   &                &                        &         &  &  &  &                &                  &         \\
                   & \multicolumn{3}{l}{$t_\beta$        -- Task-specific Saliency}   &  &  &  & \multicolumn{3}{l}{$t_\beta$        -- General Saliency}        \\
\textit{Fixed Effects}      & $\beta$           & $SE$                     & $t$      &  &  &  & $\beta$           & $SE$               & $t$      \\
\cline{2-4} \cline{8-10}
Intercept          & 73.913         & 0.569                  & 129.87  &  &  &  & 75.114         & 0.575            & 130.71  \\
FGC                &  0.996         & 0.904                  &   1.10  &  &  &  & -1.094         & 0.889            &  -1.23   \\
FC                 &  3.713         & 0.879                  &   4.22  &  &  &  &  3.835         & 0.895            &   4.29    \\
FG                 & -0.881         & 1.065                  &  -0.83  &  &  &  & -0.558         & 1.086            &  -0.51   \\
\textit{Random Effects}    & $Var$            & $SD$                   &         &  &  &  & $Var$            & $SD$               &         \\
\cline{2-4} \cline{8-10}
Subject: Intercept & 10.3581        & 3.218                  &         &  &  &  & 10.5603        & 3.250            &         \\
Subject: FGC       & 26.1320        & 5.112                  &         &  &  &  & 25.2455        & 5.024            &         \\
Subject: FC        & 24.6481        & 4.965                  &         &  &  &  & 25.5472        & 5.054            &         \\
Subject: FG        & 36.2232        & 6.019                  &         &  &  &  & 37.7145        & 6.141            &         \\
Residual           & 13.7678        & 3.711                  &         &  &  &  & 13.3568        & 3.655            &         \\
                   &                &                        &         &  &  &  &                &                  &         \\
                   & \multicolumn{3}{l}{$q$           -- Task-specific Saliency}      &  &  &  & \multicolumn{3}{l}{$q$           -- General Saliency}        \\
\textit{Fixed Effects}      & $\beta$           & $SE$                     & $t$      &  &  &  & $\beta$           & $SE$               & $t$      \\
\cline{2-4} \cline{8-10}
Intercept          &  2.598         & 0.028                  & 93.99   &  &  &  &  2.574         & 0.028            & 91.44   \\
FGC                &  0.017         & 0.038                  &  0.45   &  &  &  &  0.074         & 0.039            &  1.89   \\
FC                 & -0.178         & 0.041                  & -4.39   &  &  &  & -0.181         & 0.042            & -4.36   \\
FG                 &  0.053         & 0.054                  &  0.97   &  &  &  &  0.046         & 0.056            &  0.82   \\
\textit{Random Effects}    & $Var$            & $SD$                   &         &  &  &  & $Var$            & $SD$               &         \\
\cline{2-4} \cline{8-10}
Subject: Intercept & 0.0244         & 0.156                  &         &  &  &  & 0.0253         & 0.159            &         \\
Subject: FGC       & 0.0466         & 0.216                  &         &  &  &  & 0.0484         & 0.220            &         \\
Subject: FC        & 0.0525         & 0.229                  &         &  &  &  & 0.0552         & 0.235            &         \\
Subject: FG        & 0.0948         & 0.308                  &         &  &  &  & 0.0992         & 0.315            &         \\
Residual           & 0.0262         & 0.162                  &         &  &  &  & 0.0261         & 0.161            &        \\
                   &                &                        &         &  &  &  &                &                  &         \\

\bottomrule
\end{tabular}

%% file: TabF3.txt
\begin{tabular}{l@{ - }l r@{.}l r@{.}l r@{.}l c r@{.}l r@{.}l r@{.}l c r@{.}l r@{.}l r@{.}l}
\noalign{\vskip0.3cm}
\toprule
\multicolumn{22}{l}{} \tabularnewline

\multicolumn{2}{l}{\textit{}} & \multicolumn{6}{l}{Experimental data} & & \multicolumn{6}{l}{Simulated data} &  & \multicolumn{6}{l}{Simulated data}\tabularnewline
\multicolumn{2}{l}{\textit{}} & \multicolumn{6}{l}{} & & \multicolumn{6}{l}{General Saliency} &  & \multicolumn{6}{l}{Task-specific Saliency}\tabularnewline
\hline
\multicolumn{22}{l}{} \tabularnewline
\multicolumn{2}{l}{} & \multicolumn{2}{c}{$\beta$} & \multicolumn{2}{c}{$SE$} & \multicolumn{2}{c}{$t$} & & \multicolumn{2}{c}{$\beta$} & \multicolumn{2}{c}{$SE$} & \multicolumn{2}{c}{$t$} & &  \multicolumn{2}{c}{$\beta$} & \multicolumn{2}{c}{$SE$} & \multicolumn{2}{c}{$t$}\tabularnewline
\cline{3-8} \cline{10-15} \cline{17-22}

\multicolumn{2}{l}{\textit{Fixed Effects}} & \multicolumn{2}{c}{} & \multicolumn{2}{c}{} & \multicolumn{2}{c}{}\tabularnewline

Guess         & Count                   &  0&10  & 0&009    & 10&66               & &    0&05 & 0&009 &  4&89                      & &  0&07 & 0&009 &  7&23  \tabularnewline
CountAnimals  & CountPeople             &  0&07  & 0&013    &  5&27               & &    0&08 & 0&013 &  5&99                      & &  0&05 & 0&013 &  3&84  \tabularnewline
GuessTime     & GuessCountry            &  0&00  & 0&014    & -0&05               & &    0&00 & 0&014 &  0&36                      & &  0&01 & 0&014 &  1&09  \tabularnewline

\multicolumn{22}{l}{} \tabularnewline
\multicolumn{2}{l}{} & \multicolumn{2}{c}{$Var$} & \multicolumn{2}{c}{$SD$} & \multicolumn{2}{c}{} &  & \multicolumn{2}{c}{$Var$} & \multicolumn{2}{c}{$SD$} & \multicolumn{2}{c}{} & & \multicolumn{2}{c}{$Var$} & \multicolumn{2}{c}{$SD$} & \multicolumn{2}{c}{}\tabularnewline
\cline{3-8} \cline{10-15} \cline{17-22}
\multicolumn{2}{l}{\textit{Random Effects}} & \multicolumn{2}{c}{} & \multicolumn{2}{c}{} & \multicolumn{2}{c}{}\tabularnewline

\multicolumn{2}{l}{Subject: Intercept} & 0&0072  &  0&085 &\multicolumn{2}{l}{}   & &    0&0036 & 0&060 &\multicolumn{2}{l}{}     & & 0&0053 & 0&073 & \multicolumn{2}{l}{}    \tabularnewline
\multicolumn{2}{l}{Image: Intercept}   & 0&0167  &  0&129 &\multicolumn{2}{l}{}   & &    0&0016 & 0&040 &\multicolumn{2}{l}{}     & & 0&0015 & 0&039 & \multicolumn{2}{l}{}    \tabularnewline
\multicolumn{2}{l}{Residual}           & 0&7338  &  0&857 &\multicolumn{2}{l}{}   & &    0&7348 & 0&857 &\multicolumn{2}{l}{}     & & 0&7225 & 0&850 & \multicolumn{2}{l}{}    \tabularnewline

\multicolumn{22}{l}{} \tabularnewline
\cline{3-8} \cline{10-15} \cline{17-22}
\multicolumn{22}{l}{} \tabularnewline
\multicolumn{2}{l}{ Number of obs}   & \multicolumn{2}{l}{34188}   &  \multicolumn{2}{l}{}   & \multicolumn{2}{l}{} & &  \multicolumn{2}{l}{33941}   &  \multicolumn{2}{l}{}   & \multicolumn{2}{l}{} & & \multicolumn{2}{l}{34089}   &  \multicolumn{2}{l}{}   & \multicolumn{2}{l}{}     \tabularnewline
\multicolumn{2}{l}{Number of groups} & \multicolumn{2}{l}{subject} &  \multicolumn{2}{l}{32} & \multicolumn{2}{l}{} & &  \multicolumn{2}{l}{subject} &  \multicolumn{2}{l}{32} & \multicolumn{2}{l}{} & & \multicolumn{2}{l}{subject} &  \multicolumn{2}{l}{32} & \multicolumn{2}{l}{}     \tabularnewline
\multicolumn{2}{l}{}                 & \multicolumn{2}{l}{image}   &  \multicolumn{2}{l}{30} & \multicolumn{2}{l}{} & &  \multicolumn{2}{l}{image}   &  \multicolumn{2}{l}{30} & \multicolumn{2}{l}{} & & \multicolumn{2}{l}{image}   &  \multicolumn{2}{l}{30} & \multicolumn{2}{l}{}     \tabularnewline
\multicolumn{22}{l}{} \tabularnewline
\bottomrule
\end{tabular}

%% file: TabF4.txt
\begin{tabular}{l@{ - }l r@{.}l r@{.}l r@{.}l c r@{.}l r@{.}l r@{.}l c r@{.}l r@{.}l r@{.}l}
\noalign{\vskip0.3cm}

\toprule
\multicolumn{22}{l}{} \tabularnewline

\multicolumn{2}{l}{\textit{}} & \multicolumn{6}{l}{Experimental data} & & \multicolumn{6}{l}{Simulated data} &  & \multicolumn{6}{l}{Simulated data}\tabularnewline
\multicolumn{2}{l}{} & \multicolumn{6}{l}{} & & \multicolumn{6}{l}{General Saliency} &  & \multicolumn{6}{l}{Task-specific Saliency}\tabularnewline
\hline
\multicolumn{22}{l}{} \tabularnewline
\multicolumn{2}{l}{} & \multicolumn{2}{c}{$\beta$} & \multicolumn{2}{c}{$SE$} & \multicolumn{2}{c}{$t$} & & \multicolumn{2}{c}{$\beta$} & \multicolumn{2}{c}{$SE$} & \multicolumn{2}{c}{$t$} & &  \multicolumn{2}{c}{$\beta$} & \multicolumn{2}{c}{$SE$} & \multicolumn{2}{c}{$t$}\tabularnewline
\cline{3-8} \cline{10-15} \cline{17-22}

\multicolumn{2}{l}{\textit{Fixed Effects}} & \multicolumn{2}{c}{} & \multicolumn{2}{c}{} & \multicolumn{2}{c}{}\tabularnewline

Guess         & Count        &  0&02 & 0&005 &  4&73   & &     0&03 & 0&005 &  5&05     & &  0&04 & 0&005 &  8&37  \tabularnewline
CountAnimals  & CountPeople  & -0&05 & 0&007 & -6&88   & &    -0&04 & 0&007 & -5&77     & & -0&02 & 0&007 & -3&17  \tabularnewline
GuessTime     & GuessCountry &  0&02 & 0&007 &  3&39   & &     0&01 & 0&008 &  0&75     & &  0&03 & 0&008 &  3&87  \tabularnewline

\multicolumn{22}{l}{} \tabularnewline
\multicolumn{2}{l}{} & \multicolumn{2}{c}{$Var$} & \multicolumn{2}{c}{$SD$} & \multicolumn{2}{c}{} &  & \multicolumn{2}{c}{$Var$} & \multicolumn{2}{c}{$SD$} & \multicolumn{2}{c}{} & & \multicolumn{2}{c}{$Var$} & \multicolumn{2}{c}{$SD$} & \multicolumn{2}{c}{}\tabularnewline
\cline{3-8} \cline{10-15} \cline{17-22}
\multicolumn{2}{l}{\textit{Random Effects}} & \multicolumn{2}{c}{} & \multicolumn{2}{c}{} & \multicolumn{2}{c}{}\tabularnewline

\multicolumn{2}{l}{Subject: Intercept} & 0&0072  &  0&085 &\multicolumn{2}{l}{}   & &    0&0081  & 0&090 &\multicolumn{2}{l}{}     & & 0&0077 & 0&088 & \multicolumn{2}{l}{}    \tabularnewline
\multicolumn{2}{l}{Image: Intercept}   & 0&0022  &  0&047 &\multicolumn{2}{l}{}   & &    0&0003  & 0&018 &\multicolumn{2}{l}{}     & & 0&0005 & 0&022 & \multicolumn{2}{l}{}    \tabularnewline
\multicolumn{2}{l}{Residual}           & 0&2122  &  0&461 &\multicolumn{2}{l}{}   & &    0&2364  & 0&486 &\multicolumn{2}{l}{}     & & 0&2344 & 0&484 & \multicolumn{2}{l}{}    \tabularnewline

\multicolumn{22}{l}{} \tabularnewline
\cline{3-8} \cline{10-15} \cline{17-22}
\multicolumn{22}{l}{} \tabularnewline

\multicolumn{2}{l}{ Number of obs}   & \multicolumn{2}{l}{34873}   &  \multicolumn{2}{l}{}   & \multicolumn{2}{l}{} & &  \multicolumn{2}{l}{34873}   &  \multicolumn{2}{l}{}   & \multicolumn{2}{l}{} & & \multicolumn{2}{l}{34873}   &  \multicolumn{2}{l}{}   & \multicolumn{2}{l}{}     \tabularnewline
\multicolumn{2}{l}{Number of groups} & \multicolumn{2}{l}{subject} &  \multicolumn{2}{l}{32} & \multicolumn{2}{l}{} & &  \multicolumn{2}{l}{subject} &  \multicolumn{2}{l}{32} & \multicolumn{2}{l}{} & & \multicolumn{2}{l}{subject} &  \multicolumn{2}{l}{32} & \multicolumn{2}{l}{}     \tabularnewline
\multicolumn{2}{l}{}                 & \multicolumn{2}{l}{image}   &  \multicolumn{2}{l}{30} & \multicolumn{2}{l}{} & &  \multicolumn{2}{l}{image}   &  \multicolumn{2}{l}{30} & \multicolumn{2}{l}{} & & \multicolumn{2}{l}{image}   &  \multicolumn{2}{l}{30} & \multicolumn{2}{l}{}     \tabularnewline

\multicolumn{22}{l}{} \tabularnewline
\bottomrule 
\end{tabular}

%% file: TabF5.txt
\begin{tabular}{lllllll}
\noalign{\vskip0.3cm}
\toprule
\multicolumn{7}{l}{} \\
\multicolumn{2}{l}{Dependent variable} &  \multicolumn{2}{l}{Fixed effect part} & \multicolumn{2}{l}{Random effect part}  \\
\hline
\multicolumn{7}{l}{} \\

\multicolumn{7}{l}{\textit{Structure of LMM with custom contrast -- Comparison of the model likelihood gain in Figure \ref{fig:like_comp}b*}}\\
Model likelihood gain &$\sim$& 1 + FModel + FSal + FInter & + & (1 | subject) & + & (1 | image) \\
\multicolumn{7}{l}{} \\
\multicolumn{7}{l}{\textit{Structure of LMM with treatment constrast -- Comparison of the model likelihood gain in Figure \ref{fig:like_comp}b*}}\\
Model likelihood gain &$\sim$&  T1 + T2 + T3 + T4 & + & (1 | subject) & + &  (1 | image) \\
\multicolumn{7}{l}{} \\
\multicolumn{7}{l}{\textit{Structure of the 18 LMMs in Figure \ref{fig:nsal_compare}}}\\
Model Parameter &$\sim$&  1 + FGC + FC + FG  & + & \multicolumn{3}{l}{(1 + FGC + FC + FG || subject)}  \\
\multicolumn{7}{l}{} \\
\multicolumn{7}{l}{\textit{Structure of the three LMMs of log saccade amplitudes in Figure \ref{fig:sacstats}a*}}\\
Log Saccade Amplitudes &$\sim$&  1 + FGC + FC + FG  & + & (1 | subject) & + & (1 | image) \\
\multicolumn{7}{l}{} \\
\multicolumn{7}{l}{\textit{Structure of the three LMMs of log fixation duration in Figure \ref{fig:sacstats}b*}}\\
Log Fixation Duration &$\sim$&  1 + FGC + FC + FG  & + & (1 | subject) & + & (1 | image) \\
\multicolumn{7}{l}{} \\
\bottomrule

\end{tabular}\\